# (V4) Increasing Exclusion: The Pauli Exclusion Principle and Energy Conservation for Bound Fermions Are Mutually Exclusive

By Jonathan Phillips

Distinguished National Lab Professor, University of New Mexico

ABSTRACT

A review of those forms of standard quantum mechanics that include the Pauli Exclusion Principle as it is applied to atomic species, (that is versions of quantum that are multi-electron and multi-orbital) shows they are not consistent with energy conservation. Particular focus is given to helium in which it is shown that energy conservation is not consistent with current models. If the two electrons in the ground state, per current theory, are at the same energy as the ionization energy, it is demonstrated that according to the standard theory approximately 30 eV are lost during ionization, or alternatively, about 30 eV of energy are created during ionization/electron attachment. The same issue of energy loss during 'relaxation' of energy levels following ionization is shown to exist for all atomic species, thus demonstrating that the Pauli Exclusion Principle (PEP) and energy conservation are not consistent for any atomic species for current forms of distinguishable electron forms of quantum theory. Only that form of quantum that has a single orbital, and for which only one ionization energy can be computed (that is the original Schrödinger form), is consistent with an energy balance. However, this form is not consistent with the most common spectroscopy results, and, it is shown that the PEP has no meaning in this form of quantum theory. In contrast, a new model of quantum mechanics, Classical Quantum Mechanics (CQM), invented by R. Mills, is shown, following modification, to be consistent with all spectroscopy and energy conservation for bound electron systems. This new model is based on the validity of Maxwell's Equations and Newton's Laws at all scales. Detailed, and remarkably simple, computations for determining the 'ground state' energy levels in one and two electron systems using CQM are presented. Excellent agreement with data is found.

INTRODUCTION

   This article is intended to support two hypotheses. First, that standard quantum mechanics is inconsistent with known ionization energies. The ionization energies of the simplest multi-electron system, helium, are used to illustrate the point. In particular, we will endeavor to show that all forms (three classes are identified) of standard quantum fail, because the proposed 'relaxation' of electron energy levels, an inherent consequence of the PEP, in atoms following ionization is inconsistent with the conservation of energy. To emphasize: the failing is an inevitable consequence of assuming the Pauli Exclusion Principle (PEP) as generally understood. (The PEP as generally understood: No two electrons/fermions can have the same quantum numbers, and hence an orbital is 'full' when two electrons, identical other than having different spin directions, are in that orbital.) Second, we will show that the description of two electron systems based on a new paradigm of quantum mechanics, Classical Quantum Mechanics (CQM), developed by Randell Mills (1), but with a simple modification introduced herein, is consistent with energy conservation. Hence, it is concluded that modified CQM is a superior theory for the range of bound electrons. Extension to free electrons is left for future essays.
   In order to provide support for the above dramatic hypotheses, we include sections normally not found in scientific articles. First, we find it expedient to include a description of the scientific process as it applies to physical sciences. Second, we find it absolutely necessary to offer a new means to classify standard quantum theory. Specifically, a novel classification scheme is presented to help 'untangle' the net of quantum mechanics theories. It is argued there exist two distinct 'standard' quantum mechanics, the one 'expository' or 'descriptive' version employed by chemists and the quantitative version. Finally, there are the theories 'between' which are shown not to be true quantum in their very nature, indeed, they do not even employ Hamiltonians. These various versions of quantum are not compatible.
        Once we show that 'quantum' is not even a coherent single theory, and that the inclusion of the PEP in any quantum mechanics model makes the model inconsistent with energy conservation, we argue it is scientifically rational, in fact scientifically obligatory, to discuss alternatives to standard quantum theories. Only at this point in the presentation do we address the second hypothesis of the paper. To wit: It is shown that a new theory of quantum mechanics based solely on classical physics, particularly Maxwell's equations and Newton's Laws, and arguably one postulate regarding the strength of the interaction between two bound electrons, does give a totally satisfactory, algebraic, closed form, model of helium photo ionization with no variable parameters. Not only is it completely quantitatively consistent with all observations, but unlike standard quantum mechanics, the model is simple, self-consistent and consistent with energy conservation.

SCIENTIFIC PROCESS

In brief, physical science is a set of processes designed to produce an objective and predictive description of nature.  Relevant to the present discussions are three types of scientific processes.  One set of processes is designed to measure and record phenomenon in nature. In physics, observations considered scientifically objective share certain characteristics, including being repeatable and quantifiable.  Also, through comparison/contrast with general observed patterns of nature, objective observations are categorized.  For example, photo ionization of helium is one of the most studied processes in atomic physics. It has been repeated in countless labs and the quantities associated with the process have been verified repeatedly (2-7).  Thus, the measured aspects of the process are 'scientifically objective', and it is clear the process is classified as one of a set of objectively observed processes, all similar in nature, known as photo ionization.  The photo ionization of helium is a good example of a scientifically objective observation as it is clearly a highly repeatable process that has been precisely quantified, and belongs to a larger set of related processes, similarly repeatable and quantified.

A second process of physical science is finding 'rules', 'descriptions', and in physics in particular, quantitative THEORIES.  These theories are valuable in that they can be used to predict behaviors quantitatively. A scientifically acceptable theory must be consistent with all objective scientific observations that are known for the proposed range of validity of the theory.  Also, scientific theories should be self-consistent, and consistent with theories that are known to be consistent with all observables over the proposed range of validity (8).  What is a 'proposed range of validity'?  The best example is the fact that, standard quantum theory is purportedly valid at all size scales, but classical physical theories including Maxwell's Equations and Newton's Laws are only believed to be valid for descriptions of physical processes at scales 'greater than h-bar' (Correspondence Principle).

Theories come and theories go.  There is no heuristic for inventing theories.  The best seem to come as epiphanies and involve simple ideas.  The core equation in Einstein's special relativity revealed itself to him in a dream.  In contrast there is a simple method for eliminating theories:  scientific testing, the third category of scientific process relevant to this paper.   A theory that fails scientific testing must be eliminated.  Scientific testing is conceptually simple:  A theory cannot be proved, but it can be disproved.  Indeed, if a theory is shown to be inconsistent with just ONE objective scientific observation, either qualitatively or quantitatively, it is disproved.  At a minimum, a theory shown inconsistent with a single objective scientific observation is no longer valid, or at least not valid over some range (8). Thus, for example, if quantum theory can be shown to be inconsistent with any objective scientific observation, regarding valid scientific data for bound electrons, it is a rational, scientific, activity to seek new theories for bound electrons.

The above section/lecture should not be necessary.  The fundamentals of the scientific process should be broadly understood.  Unfortunately, the author believes many in the community are not aware that no theory is proven, hence questioning any theory is a scientifically sound activity, whereas arguing against even the consideration of alternative theories, as many scientists have in regard to quantum theory, is unscientific and irrational. In fact, regarding the current issue, alternative quantum theories, there is a considerable history of efforts to create a physical model of the bound electrons by distinguished physicists (9-12).  Although these attempts to create a planetary/solar system model of electrons/atoms ultimately failed, generally because they could not predict the correct g-factor (13), they indicate that many scientists have always understood that the current quantum model is far from satisfactory and that alternatives should be considered.

REVIEW OF STANDARD QUANTUM THEORY APPLIED TO BOUND FERMIONS

It is important in the context of the core of this paper to recognize that prior to the acceptance of the quantum model (QM) in its earliest form about 80 years ago, physicists believed that Newton's Laws and Maxwell's Equations applied at all size scales. However, no physicist succeeded in finding a means to explain the 'quantum' nature of stable energy levels in atoms, discovered via spectroscopy (e.g. Balmer series), using these laws (13). Hence, a new theory, herein called standard quantitative quantum mechanics (SQQM), was born that produced results quantitatively consistent with the quantum behavior of hydrogen. In time it was realized that this new theory, SQQM, was totally inconsistent with classical physics laws at size scales 'less than h-bar'. For example, according to SQQM bound electrons, in single electron systems, are 'point particles' that jump randomly, obeying only a statistical description of their position- not forces, not conservation of momentum, etc., hence in violation of Newton's Laws. Another clear example of the violation of Newton's Laws: In those cases in which the point electron in a one electron system is far from the nucleus, allowed and expected according to the probability distribution used to describe the electron, it has nearly zero potential energy. At those positions in order to preserve the total binding energy, that is the sum of the potential and kinetic energy, the electron is presumed to have negative kinetic energy. This is clearly not consistent with Newtonian mechanics, even if it is argued these states only last a tiny fraction of a second.

The violations of Maxwell's equations are equally clear. The random travels of the electron in the single electron system anticipated on the basis of the probabilistic model, result in instantaneous jumps into/out of the nucleus. In a classical sense this is accelerated motion, which by Maxwell's equations should require the emission (jump toward nucleus) or absorption (jump away from nucleus) of energy. Yet according to SQQM these processes occur without the emission or adsorption of energy, hence SQQM is a clear violation of Maxwell's equations.

To make SQQM match the older classical theories the community, under the guidance of Bohr, adopted the notion of the 'Correspondence Principle' more than 70 years ago. At some size scale, around h-bar, the two theories give identical predictions. Still, there are many who never were satisfied with this explanation. Einstein was very unhappy with the statistical description of the behavior of sub-atomic particles given in quantum theory, as succinctly put in his now famous comment, 'God does not play dice with the universe.'

The above issues are well rehearsed even in basic quantum theory courses. However, the problems with SQQM are far deeper for multi-electron systems, e.g. helium, than they are for single electron systems. In particular: the multi-electron wave function multiplied by its complex conjugate, a single scaler value at every point in 3N (N is the number of electrons) phase-space, *has no physical meaning*. This is discussed in more detail later.

Another issue is the notion that anything regarding SQQM is proven. Indeed, is the, above described, division of physics at a size scale of h-bar proven? No. A theory shown to be consistent with all known phenomenon is still not proved, just shown to be 'likely'. After all, science is empirical, hence not susceptible to absolute proof. Although proving a theory is difficult, disproving one is simple. Indeed, all that is required to disprove or at least limit the applicable range of a theory is to show that the theory fails to accurately describe a <u>single</u> 'objective scientific observation'. Indeed, this is the rationale given for restricting the domain of applicability of Newton's Laws and Maxwell's Equations to scales 'greater than h-bar'.

It is still possible that Newton's Laws or Maxwell's equations do apply to matter on all scales. It has not been proven that they do not, all we know is that earlier efforts failed! Thus, the argument at the core of this manuscript, that a single epiphany regarding the correct geometrical description of elementary particles 'allows' correct (i.e. matches objective scientific observations) values to be obtained using classical laws of physics on all scales, cannot be summarily rejected as 'unscientific'. If such a claim is made, it must be tested. In our case we provide a first test of the CQM physical description of bound electrons. We show that the new 'physical' model of the bound electron is quantitatively consistent with all presently known objective scientific phenomena. Will the physics community recognize the primacy of the scientific process and not be guided by the weight of tradition? Will they even consider questioning SQQM?

The sections below amplify many of the themes introduced above. This is done with respect to the three major classes of standard quantum, only one of which can be the correct quantum. Discussions of standard quantum, including defenses of standard quantum, must necessarily begin with the identification of the class of quantum theory at issue. That is, those arguing 'for' quantum theory must select, and identify, only one of the three theories below as the 'quantum' they are defending. These three theories, it is argued below, are distinct and incompatible.

THE INTUITIVE VERSION: DQM

One major difficulty in a critique of quantum mechanics is the difficulty of 'pinning down' the core theory. There are actually at least three categories of 'parallel' quantum theories. The most intuitive is herein called 'Descriptive Quantum Mechanics' (DQM). It is often erroneously conflated with SQQM. In fact, DQM is the theory generally taught to scientists and engineers. In the DQM theory, electrons are quasi-independent particles that independently occupy (hydrogen like) 'orbitals' and have energies/spins and other properties associated with the orbital they occupy. They can move between orbitals without significantly disturbing electrons in other orbitals. Each electron in a multi-electron system has a particular energy, associated with the quantum level it occupies. The total energy of the electron system, say helium, is the sum of the energies of the individual electrons. The total spin of the system is the sum of the 'spins' of the individual electrons. Thus, for example, helium in the ground state has two identical electrons at approximately -25 eV each. One important feature of DQM is that in the language of DQM the PEP actually has meaning. (In contrast, in the mathematical language of SQQM the PEP cannot be expressed in a meaningful manner. More below.) A second positive feature of DQM is that it permits a simple means of keeping track of angular momentum and thus can account for selection rules on the basis of conservation of momentum, something alternative quantum formulations (below) cannot. In any event, we show below that in DQM either PEP or energy conservation must be incorrect. We choose to believe it is the former.

DQM also can explain phenomena such as XPS data from multi-electron atoms in which it is clear that there are a multitude of ionization energies. That is, it is recognized that both 'interior electrons' and the highest energy electrons occupy stable orbitals with precise energy levels. Since its description of an atom indicates that electrons occupy specific energy levels, it is consistent with the supposition that atoms at lower energies (inner orbitals) can be 'knocked-out' of the atom (ionized) with a bolt of the correct energy. A cascade of radiation (either photons or electrons) with precise energies will follow as the outer orbital electrons in their independent orbits 'fall in energy' to fill the lower energy 'holes' created by the ionization of inner orbital electrons.

Is DQM a valid physics theory (8)? As shown below, DQM cannot even produce an energy balance consistent with photo ionization of helium, thus clearly DQM cannot be considered a valid scientific model. (Later the same question will be asked of SQQM: Is SQQM a valid physics theory? The answer is also no! It is true that in contrast to DQM, SQQM can produce a valid energy balance for the photo ionization of helium, but it will be shown it is completely inconsistent with other clearly measured phenomenon such as precise ionization energies for 'inner' electrons. )

<u>Deposing DQM</u>- Postulate: If quantum theory cannot explain photo ionization of helium, then it is not a valid scientific model for any bound electrons. Below it is shown that the DQM model cannot explain either photo ionization of helium or related phenomenon, for example electron attachment to He+.

The DQM model of the photo ionization process for atoms is that a photon with sufficient energy is absorbed by a bound (atomic) electron and hence 'frees' the electron. For example, in the case of hydrogen it is well established that a photon with a minimum energy of 13.6 eV is required to 'free' an electron, that is bring it to a state in which it no longer feels any force from its original nucleus. The fact that SQQM predicts a binding energy for the sole hydrogen electron exactly equal to the measured ionization energy is one of the great triumphs of SQQM.

A second aspect of the model of relevance is the postulate that for all two electron atoms in their ground state, the two electrons are identical in all respects, except for their spin direction/spin quantum number. In particular, the two electrons have identical energy. For helium the energy of both electrons in the DQM model is approximately -24.5 eV, because, that is the energy required for ionization. The energy assignment of -24.5 is the only rational interpretation of the 'energy level diagram' repeatedly employed in descriptions of DQM. In the DQM model the magnitude of the ionization energy and the magnitude of the energy of the electron that is being ionized are equal, just as it is the case for one electron systems. That is, the value -24.5 eV does not come necessarily from a direct quantitative theory (and DQM as defined here is not quantitative), but rather from objective scientific observation (14) of the ionization process (See Figure 1).

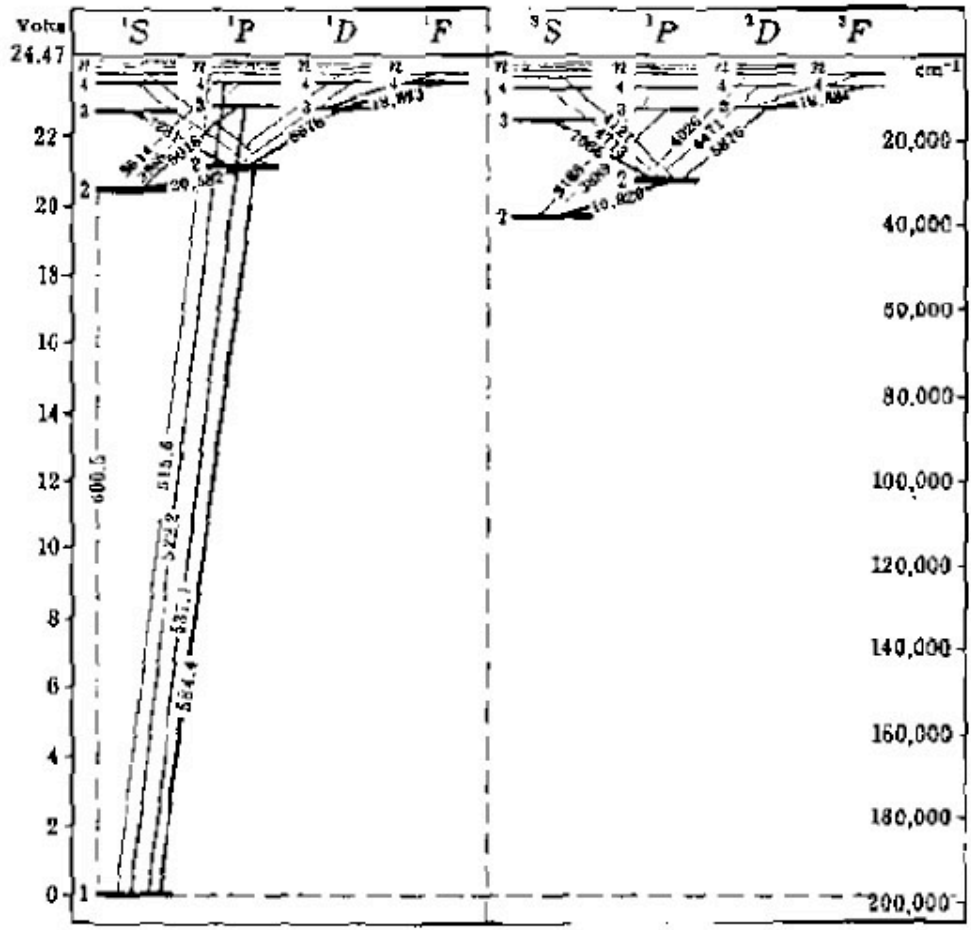

**Fig. 1**- *Classic Energy Level Diagram for Helium.* If both electrons in the ground state are not at -24.5 eV, what are their energies? (From Ref. 16)

A third aspect of DQM of relevance is that after/during ionization the remaining electron in He+ 'falls' to a lower energy state. The measured value of ionization for the electron in He+ is -54.4 eV, a value in close agreement with the ionization calculated using Schrodinger's equation for the ground state of a one electron atom with a two proton nucleus. Although it is not generally explicitly stated, in virtually any standard text (15-17), it is an inescapable implication of the model, and the only interpretation consistent with energy level diagrams (see Figure 1) and the PEP. Thus, the electron that remains after He is ionized 'falls', according to this theory, from approximately -24.7, to -54.4 eV

The fourth and final aspect of DQM of relevance is the inclusion in the theory of the Newtonian law of energy conservation. There is no 'scale' at which this general principle does not apply.

The reader, armed with the four 'aspects' of DQM discussed above can demonstrate to himself the fact that DQM cannot provide a satisfactory explanation for the simple process of photionization of helium. Thus, one concludes that the nice physical picture of the atom at the heart of DQM is not valid, and hence DQM is not a scientific theory. Specifically, the reader can show that DQM is not valid simply by consideration of the following questions. Remarkably, there is no need for math any more complex than basic algebra to disprove DQM.

1. During/after ionization the electron that is not ionized falls in energy from -24.5 to -54.4 eV. Where do the nearly 30 eV lost by the remaining bound electron in this process go? Since chemical reactions are basically the movement of electrons between states, this conundrum can be readily written in the symbolism of chemical reactions:

He (-49.0 eV) + hν (+24.5 eV) → He$^+$(-54.4 eV) + e$^-$ (0 eV); ΔHrxn= -29.9 eV   (1)

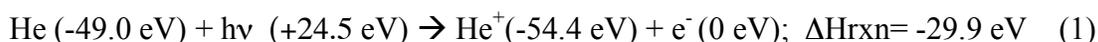

Where the energy of the species, assuming the Pauli Exclusion Principle is correct, are provided. (Incidentally, the identification of the measured ionization energy, 24.5 eV for helium, and 'energy level' are universal for this form of quantum.) The nearly 30 eV unaccounted for is an enormous energy. In contrast, the re-organization of hydrogen and oxygen to make water produces about 1.2 eV /H atom, and this is associated with a lot of sensible heat. (One plank of the U.S. national energy plan is to capture this very significant energy directly as electricity.) Moreover, it is clear the 1.2 eV is simply the net released by electrons moving from one set of stable orbitals to a new set of orbitals of lower energy that become available for occupancy upon the formation of water molecules. In precisely the same fashion the 29.9 eV of energy released (according to DQM), per Equation 1, represents the net change in the stable energy levels of the electrons in helium during ionization.

2. During electron attachment to He+ to re-create atomic helium, where does the energy come from to raise the electron energy from -54.4 to -24.5 eV? In fact, it is an objective scientific observation that during electron attachment to helium almost 24.5 eV of energy is *released* as a photon. The release of energy during the process of electron attachment to He+ is in fact required to maintain 'microscopic reversibility'.

He$^+$ (-54.4 eV) + e$^-$ (0 eV) → He (-49.4 eV) + hν (+24.7 eV); ΔHrxn= +29.7 eV   (2)

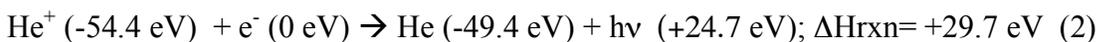

3. Why is electron attachment spontaneous if the total energy of two bound electrons in atomic helium is lower than that of ionized helium? The total energy of the electrons in atomic helium according to DQM theory (2*-24.5) is -49.0 eV. In contrast, for ionized helium, the total energy of the free electron (0 eV) and the one bound electron is -54.4 eV. Clearly ionized helium is in a lower energy state than helium. Accordingly, helium should spontaneously ionize.

He (-49.0 eV) → He$^+$ (-54.4 eV) + e- (0 eV); ΔHrxn= -5.4 eV     (3)

The failure of DQM to be consistent with energy conservation is not subject to additional study. Energy and energy change are state properties. All that is needed to evaluate consistency with energy conservation are the initial and final energy values. If two objects are 'identical' or 'indistinguishable' they must have the same state properties. Study of how the process occurs cannot impact the net energy change, any more than a study of the path a balloon travels from the top of the Sandia Mts. in New Mexico (10, 600 ft) to my front yard (5200 ft.) in Rio Rancho, NM changes the fact that the balloon lost 5400' of altitude in the process.

Most physicists conflate SQQM and DQM and attempt to save DQM because they perceive a threat to SQQM in the failure of DQM. In other words, they feel that the failure of DQM is a failure of the entire quantum hypothesis, because they don't recognize the basic duality of standard quantum theory. As we shall see, they are wrong. However, it is interesting and revealing to discuss the many arguments posited to save DQM. These arguments, actually proposed 'modifications' to DQM, are considered below.

Two categories of modification are most common: First are the 'missed release' models. This category of modifications always postulates a release of energy never previously observed during photo ionization. Clearly, this is not part of any standard model. Second are the 'stored' energy models. Modifications of this type involve energy 'storage' in the He+ ion following ionization in the form of increased mass or in 'fields'. Each category of answer is examined below.

It can quickly be shown, again with simple arguments, that all 'missed release' models lead to absurd conclusions. The patent office will not issue patents for perpetual motion machines, yet the missed release model of helium photo ionization can be shown to be equivalent to perpetual motion. Consider the following thought experiment. Helium gas is placed in a chamber with perfect mirrors. A pulse of 24.7 eV light from a laser is added. This energy will ionize the helium. In consequence, in these models, there is an increase in the effective nuclear charge to a full +2 on the remaining electron. At this point, almost 30 eV are released, presumably in the form of a photon. This is the 'missed release' of energy, 'missed' as it has never been observed. Next, as per valid scientific observation (and the principle of microscopic reversibility) the free electrons re-attach to He+ ions to reform atomic helium. This process is known to release the same amount of energy as required for ionization, about 25 eV of energy! Clearly, this model again violates the energy balance. The system starts and ends in the identical form: helium. No fuel is burned, yet a net of almost 30 eV of 'free' energy is generated for each photon of 25 eV put into the system. Is helium ionization a possible solution to the energy crisis? Think again.

The proposed 'missed emission' process can be written as two steps. First, helium is ionized and emits a photon:

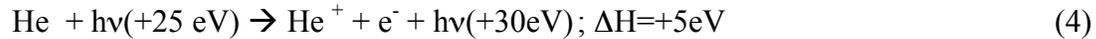

$$\text{He} + h\nu(+25\text{ eV}) \rightarrow \text{He}^+ + e^- + h\nu(+30\text{eV}); \Delta H=+5\text{eV} \qquad (4)$$

And then the helium ion recaptures an electron:

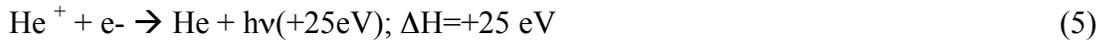

$$\text{He}^+ + e^- \rightarrow \text{He} + h\nu(+25\text{eV}); \Delta H=+25\text{ eV} \qquad (5)$$

Adding 5 and 6 together yields no change in the system (helium in, and helium out!) except for the net release of nearly 30 eV of energy!

Another of the 'missed emission' modifications of SQM is that the ionized electron and the remaining He$^+$ absorb the energy in 'recoil'. However, as in the co-ordinate system of the original atom momentum must be conserved, the electron, being almost 8000 times lighter than the ion, will have virtually all of the released energy. Is it possible the physics community never noticed 30 eV electrons produced during photo ionization? In any event, the Perpetual Motion argument applies to this model as well.

The 'stored energy' models all suffer from a lack of precision and quantification. Where is the energy stored? Is it in the form of mass? Which particle in the ion gains mass during photo ionization? Is this even possible given that the rest mass of fundamental particles is immutable? The conversion of mass to energy involved entire particles, not parts thereof. How is the mass reconverted to energy during electron attachment? If the energy is postulated to be stored in 'fields' the question is: What fields? Such a model must be quantified, or it simply has no credibility.

<u>Beyond the Literature: A Proposed Modification to DQM</u>- It is possible to imagine a modified DQM that is consistent with both energy conservation and the Pauli Exclusion Principle. Warning I: This modified version is not found in the literature (e.g. 14-17). Warning II: the modified version is difficult to test. In respect to its consistency with an energy balance the modified model is superior, however the modified model arguably contains some metaphysics as it is not possible to compare the computed electron energy values by any direct process. In particular, for these models, the measured ionization energy and energy levels of the excited states, do not correspond to measured values of ionization energy or excitation.

'Abridged' /'schematic' energy levels for all of the possible new models of helium are illustrated in Figure 2. In all three cases to make the PEP and an energy balance work at the same time, we must introduce the concept of 'boost'. This is postulated to be a process in which the energy released as the unionized electrons fall to there post-ionization levels, the so-called 'relaxation energy', help 'boost' the ionized electron out of the atom. There are three versions for helium. In the 'NON INTERACTING' model the two electrons do not interact, and the 'unexcited' electron remains at its initial energy level until ionization. The key features are i) both electrons are initially at ~-39.5 eV (agrees with total observed electron energy of -79 eV), ii) None of the energy levels of excited states are the same as in the 'Standard Model', iii) spacings between energy levels are the same as in Figure 1, iv) no energy level corresponds to -24.5 eV, v) there is a gap of about 15 eV between the highest excited state and the vacuum energy and vi) the jump to the vacuum level occurs when the 'unexcited' electron falls to its final energy of nearly -55 eV.  In the 'INITIAL DROP' model, the electrons also do not interact, but the 'unexcited' electron falls to its final energy state under any and all excitation processes. Thus the first excitation process puts the energy of the non-ionizing electron 'drop', all the way to -54.4 eV, into the excited/ionized electron.  After the first step the non-ionizing electron remains at its 'ionized' He level though all subsequent excitation or ionization processes.  In this model, the excited state energy levels of the 'excited/ionized' electron match those of the 'STANDARD MODEL' and there is no large energy gap between excited states and the vacuum.

In the 'STAGED DROP' model, it is assumed that there is electron-electron interaction energy.  The total energy of each electron is -39.5 eV, but that energy is made up of negative energy of binding with the nucleus and some positive, repulsive, 'bond' energy coming from the interaction between the electrons.  For example, there could be a (repulsive) total 'bond' energy of +10 eV, and each electron could be bonded to the nucleus with an energy of -44.5 eV.  As the two electrons are identical, it is rational to assign half of the 'bond' energy to each electron.  Thus, the net energy of each electron is -39.5, as shown. (This assignment of half the bond energy to each electron is certainly consistent with standard quantum which always assumes the existence of a repulsive force, that is positive energy, between electrons, yet assigns specific energy levels to the electrons.)  With each photon adsorption/excitation process there is a concomitant drop in the energy of the unexcited electron. Thus the 'unexcited' electron occupies an entire set of energy levels, one for each excited state. Other notable features of this model: All of the excited states are at lower energies than those given in the 'Standard Model', and all of the energy spacings between excited energy levels are larger than those of the standard DQM model.

All of the features of all of the energy level models present in Figure 2, EXCEPT THOSE OF THE STANDARD MODEL, are consistent with the PEP, and an energy balance.  That is, the energy required for ionization in the standard model is -24.5 eV, exactly the observed input energy during the first ionization of helium.   Thus, there is no accounting for the energy lost when the 'unionized' electron falls nearly 30 eV in energy following the first ionization.  In contrast, the other models all account for the energy lost by the 'unionized' electron when it looses, in all the new models, about 15 eV.  That energy, plus the input energy of  24.5 eV are required to boost an electron initially at -39.5 eV to the vacuum level. The 'Staged Drop' model is unique in that the positive repulsive interaction energy is given up in a series of steps, however, all energy is accounted for in this model.

Given present information it is impossible to tell which, if any, is correct.  It is abundantly clear that no current quantum model of the energy levels in helium agrees with any of them:  No current model starts with the two electrons in the ground state at -39.5 eV, and none provide ANY information regarding the energy state of the 'unexcited/unionized' electron during the excitation process.

# 'STANDARD MODEL'

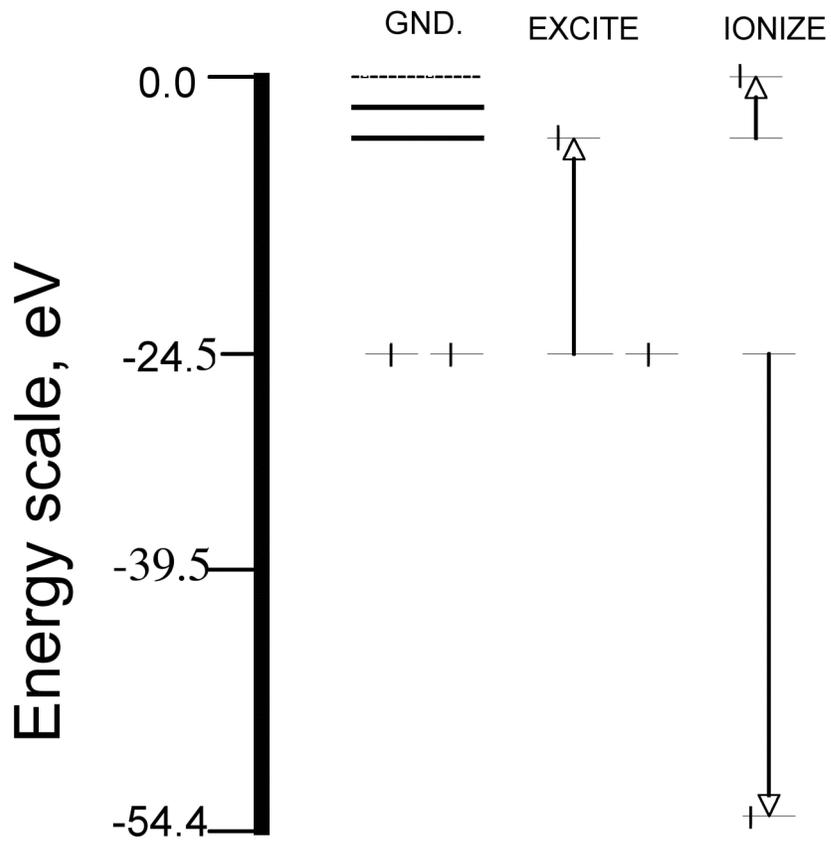

# 'NON INTERACTING MODEL'

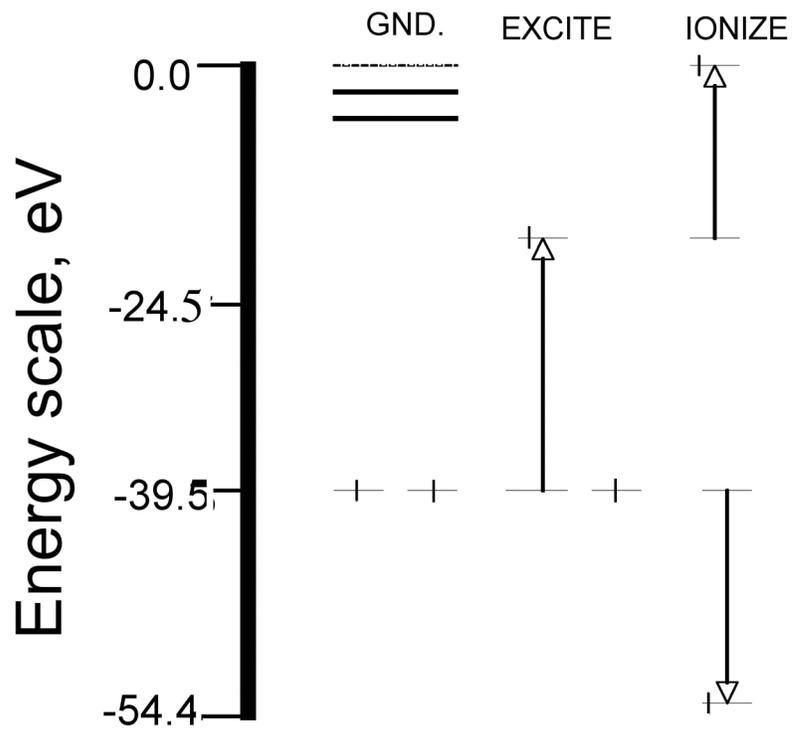

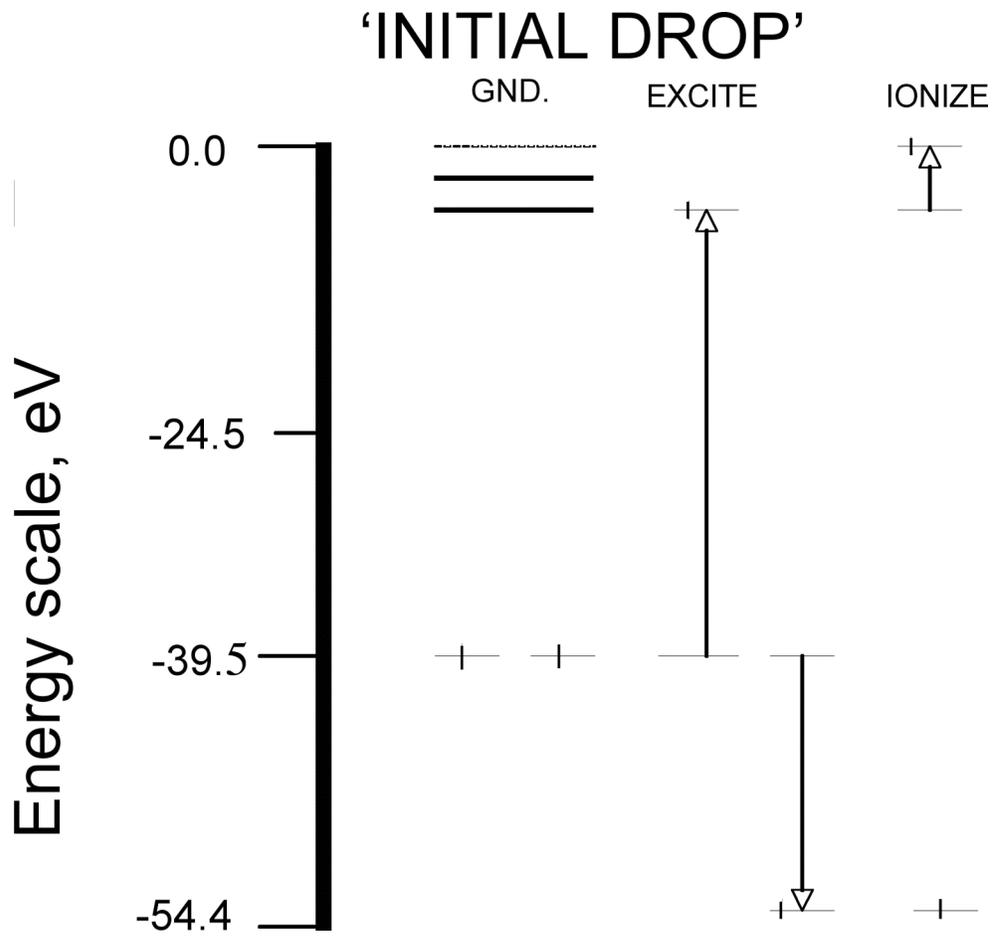
'INITIAL DROP'

# 'STAGED EXCITATION'

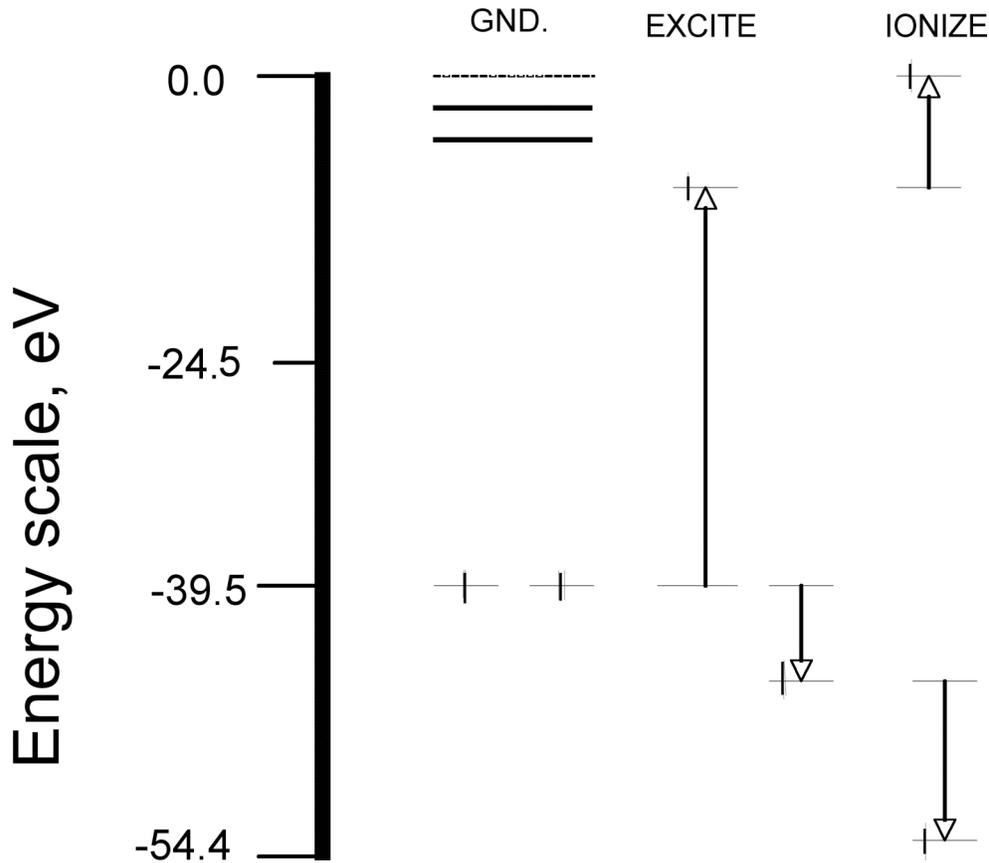

**Figure 2:** *Novel DQM Models Consistent with PEP and Energy Balance*. In all of the above novel ('Non-Interacting', 'Initial Drop', 'Staged Excitation') abridged/schematic models of helium, both electrons in their ground state are at -39.5 eV. In contrast, in the 'Standard Model', both electrons are initially at -24.7 eV, the only rational interpretation of the usual energy level diagrams (see Fig. 1 or any text) and the PEP. In all cases a two step process is illustrated. First, one electron is 'excited', and the other electron, according to the particular model, either remains at its initial energy level, or drops to a lower energy. Second, the excited electron is 'ionized' starting from its excited state. The 'unexcited' electron in all cases is found at -54.4 eV after this process. Finally, it is notable that in the Standard Model the energy for both the excitation and ionization step (a net of 24.7 eV, per experiment) come from photon adsorption. In contrast, for the other models some of the energy comes from photon adsorption (a net of 24.7 eV) and some from the drop in energy of the non-excited electron from -39.5 eV to its final value of -54.4 eV ( a net of approximately14.7 eV). The sum of these two energies equal that required to ionize an electron initially at -39.5 eV. Thus in the standard model the energy lost during the 'drop' of the unionized

electron from -24.7 to -54.4 is not accounted for.  Hence, there is approximately 30 eV of energy 'lost to the universe' during the ionization process as it is described in the Standard Model.

THE REAL DEAL:  SQQM

SQQM, is the actual quantitative quantum theory developed by Heisenberg, Schrödinger, Bohr, etc.  As applied to atoms and ions SQQM does produce a model of photo ionization of helium that is consistent with energy conservation, but it must 'give up' all the physical content of DQM, in fact all physical meaning, in order to do so. Note: SQQM is a far more sophisticated mathematical 'theory' than DQM and hence different arguments must be brought to bear against it than those employed against DQM.  Thus, the arguments given below, particularly those regarding the inherent inconsistency of the Pauli Exclusion Principle and Energy Conservation, as per the title of the manuscript, are distinct from those employed in the DQM section.

The core arguments against multi-electron SQQM are: i) the PEP is technically meaningless as applied to SQQM, ii) SQQM does not have physical meaning for multi-electron systems, iii) SQQM is mathematically inconsistent because it switches back and forth between phase space and real space, iv) SQQM is self-inconsistent because for one electron the wave function is a probability map in real space, whereas for multi-electron systems there is no real space map and no probability distribution  (acceptable theories are consistent theories), and v) SQQM arbitrarily drops some elements of classical physics (Newton's Laws, Maxwell's equations, magnetic force), keeps others (electrostatic potentials) and invents others (correlation energy).

In SQQM applied to multi-electron systems, there are no 'individual electrons'. Indeed, there is no such quantity as 'the energy of an electron' or the 'spin of an electron' or the 'angular momentum of an electron', or even the 'probability distribution of an electron'.  In SQQM one develops a single 'wave function', from a single Hamiltonian, not a number of Hamiltonians, that 'number' being set equal to the number of electrons. That would be the number required to obtain wave functions for each 'particular' electron.  The Hamiltonian is intended to express all the energies arising from forces. Surprisingly, it never includes a magnetic interaction between electrons.  For two electron systems, the Hamiltonian is written in non-operator form (18):

$$H = \frac{p_1^2}{2m} + \frac{p_2^2}{2m} - \frac{Ze^2}{r_1} - \frac{Ze^2}{r_2} + \frac{e^2}{|r_1 - r_2|} \qquad (6)$$

Where p is the momentum, Z the nuclear charge and m is the reduced mass.  The generalization to many electron systems is clear.  One adds an additional kinetic and nuclear electrostatic term for each electron, and the number of two electron repulsive terms is one less than the number of electrons.  This leads to the generation of a complex wavefunction that is most compactly expressed in the form of a Slater determinant (18), which automatically yields the anti- symmetric (independent of the labels on the electrons) form appropriate for Fermions (fractional spin particles). In particular for a two electron system the wave function is most easily expressed as a product-sum of 'single particle' wave functions and has this general form:

$$\Phi = \frac{1}{\sqrt{2}}[\phi_1(A)\phi_2(B) \pm \phi_1(B)\phi_2(A)] \tag{7}$$

Where the subscripts represent the different 'single particle' wave functions ($\phi$), and the letters in parentheses are the electron identifiers. The above is actually multiplied by a 'spin function', and the nature of the spin function determines the identity of the +/-. A minus sign is correct if the spins are parallel and positive if they are anti- parallel, to insure total antisymmetry of the wavefunction.

Note that mathematically all electrons are 'indistiguishable'. That is, if the electrons are interchanged, the wave function is not changed. The mathematically indistinguishable nature of the electrons is presumed to reflect the 'real' world. In other words, the bound electrons of an atom are sort of 'one big electron', hence describable by one big Hamiltonian.

Multi-Electron Physical Meaning- The above discussion raises a fundamental question: What is the 'meaning' of a multi-electron wave function in SQQM? The answer, NONE. That this answer is correct is found from a consideration of three issues: normalization (phase space), computation of electron-electron interaction (real space), and examination of the values of the wave function 'mapped' from phase space into real space. It is argued that once properly understood simultaneous consideration of these three issues shows that SQQM is not even a valid scientific theory, and the wave function for multi-electron systems has no physical meaning. Indeed, it is clear that even the mathematics employed in the first two standard processes are inconsistent. This violates one of the fundamental requirements of a scientific theory: it must be self consistent. Another violation of the requirement of consistency is the nature of the wave function. Single electron wave functions can be plotted in real space. For single electron wave functions each point in real space has a single value and that value has a clear (?) meaning: The probability that the electron will be at that point. As shown below, each point in real space for a multi-electron wave function has an infinite set of value. As a consequence, multi-electron wave functions cannot be graphed.

The first issue is the method employed to 'normalize' the multi-electron wave function. Of particular concern is the use of 3N (N is the number of electrons) 'phase space' to make the value of this function integrated over all 'space' equal one. Standard practice (19) is to formulate the wave function from a set of orthonormal orbital wave functions. Examination of Eq. 7 shows that any wave function composed of orthonormal component functions, multiplied by its complex conjugate, and integrated over 'phase space', has a normalized value of one. Indeed, the value for the Slater determinant antisymmetric wave function (18) for any number of electrons is one.

The reader can show that the wave function multiplied by its complex conjugate actually becomes the sum of four terms, each having four of the original one electron/hydrogen type wave functions. Two of these terms integrate to zero because the wave functions are orthonormal in both 3D phase spaces. Two of these terms integrate to one, because over both phase spaces the integrated values of the wave function and its complex conjugate are one, hence the product of the two integrations is one. It is notable that the number of dimensions required for integration/normalization increases with the number of electrons. For two electrons, two 3 dimensional phase spaces are required, for three electrons the total number of dimensions is nine, etc.

It might be reasonable to argue that the 'phase space' approach is acceptable if it were employed consistently. It is not. Indeed, the last term in the multi-electron Hamiltonian (Eq. 6) requires computing the energy of electrostatic interaction between two electrons in real space. (This in itself is odd. Given that Newton's Laws and Maxwell's equations are dismissed 'at the level of h-bar' why not electrostatic interactions? And why is there no term in the Hamiltonian for energetic interactions between the magnetic moments of electrons?) Apparently, and not for sake of consistency, it is believed that electrons cannot interact in 'phase space', they can only interact in real space!

Historically there has been a great deal of confusion on this very point. For example, we show below how two giants of quantum mechanics suggested means to understand and compute this interaction that are fundamentally flawed.

Examine the electron-electron interaction term in the two electron Hamiltonian (eq. 6), where r ($r=r_1-r_2$) is the distance between the electrons, in *real* space. Clearly, at any point in space where the two electrons are at the same position ($r_1=r_2$) there is a singularity at which point the repulsive energy term in the Hamiltonian goes to infinity:

$$E = \int \phi | \frac{1}{r} | \phi^* dr \tag{8}$$

where E in this case is the repulsive energy. Early in the development of SQQM, methods to evaluate this integral containing the very apparent singularity were offered. These explanations continue to be cited, and the results accepted. However, as we show below, they are clearly not correct. For example, Eyring *et al*. (20) suggested the following equality could be used, and would lead to a finite value of the repulsive energy

$$\frac{1}{r_{i,j}} = \sum_{n=0}^{\infty} \sum_{m=-n}^{m=+n} \frac{4\pi}{2n+1} \frac{r_<^n}{r_>^{n+1}} Y_n^m(\theta_i \phi_i) Y_n^{m*}(\theta_j \phi_j) \tag{9}$$

where $r_{ij}$ is the "distance between the two particles" and "$r_>$ be the greater of $r_i$ and $r_j$ and $r_<$ be the lesser". Developed using a 'proof' that appears questionable, Eyring's proposed substitution leads to the above expression, one in which the equal sign is clearly wrong. Indeed, consider that $r_<$ is at the origin, and $r_>$ is any point but the origin. In this case, every term in the sum contains a zero co-efficient ($r_</r_>$), thus each term is zero, and hence the sum for any $r_>$ and any angle is zero. Yet, it is clear that $1/r_{ij}$ in this case is not zero.

Pauling *et al*. (21), took an alternative approach to evaluation of the integral with the singularity. The approach is not mathematical at all. It begins with an assertion: The integral representing the electrostatic repulsion between two electrons is equivalent to, "the mutual electrostatic energy of two spherically symmetric distributions of electricity with density functions $e^{-r_1}$ and $e^{-r_2}$", and he proposes to integrate this over *real* space. No explanation, no mathematical proofs, are offered. Even if this assertion is accepted as true, it is quite clear that such charge distributions would have infinite energy as there would be non-zero overlap of charge at all points in space. The repulsive energy of any non-zero charge overlapping any other non-zero finite charge at any single point is infinite (e.g. see discussions of electron self-energy (22)). There may be a mathematical rationalization for the final expressions, however, there is clearly no physical explanation. This is simply one more example supporting the earlier offered conclusion that there is no physical meaning associated with standard quantum theory as applied to multi-electron systems.

How is the difficulty of the singularity in real space avoided? The approach generally taken is to propose that electron motion is correlated. For example, in a two electron system, if one electron is at Point R1, then the other electron has 'correlated motion' such that the probability that it is also at Point R1 is 'low' (zero?) and in fact it is 'probably' not in the general region of Point R1 at all.

Those who insist that the multi-electron wave function is 'physical' indicate that the wave function is really defined by this correlation. The wave function is a measure of the probability of a particular geometry of electrons. Thus, for a two electron system the value of the wave function at any point in space (R1) is the probability that one electron is at R1 and the second electron is at R2. This can be a small value if R2 is in the vicinity of R1, or a relatively large value if R2 is far from R1. But, there is some value for every geometry. To continue: The wave function at R1 must also be a measure of the probability that one electron is at R1 and the second electron is at R3. The wave function at this point must also be a measure of the probability that one electron is at R1 and the second at R4, etc., etc., etc. Clearly, the consequence of accepting this interpretation of the wave function is that the wave function must have an infinite set of values at every point in real space. This is because if one electron is at this point, there are an infinite number of positions that the other electron can be found, some more probable than others. There must be a probability value for each of these possibilities associated with Point R1, hence there are an infinite number of values at point R1 in real space. This is not only 'ungraphable', it is not a physically plausible interpretation.

Alternatively the two electron wave function can be described using a 'phase space' picture. For every position Electron 1 takes in its 'phase space', there is a matching probability distribution, encompassing every point in the 'phase space' of Electron 2. (Perhaps, it can be argued, every change in Electron 1 position changes the Hamiltonian of Electron 2 in its separate phase space.) Since Electron 1 can be an infinite number of positions, there are an infinite number of Electron 2 distributions in Electron 2 phase space. The real space description given before, in which each point in space has an infinite number of values, can be considered a 3D 'mapping' of 6 D phase space. What about the element lead, for which the electron 'phase space' has 246 dimensions? At a minimum, it is clear that the multi-electron wave function does not have a 'physical meaning' even remotely similar to that of the one electron wave function. SQQM is not a self-consistent theory.

Is there any other possible physical meaning to the multi-electron wave function? The first difficulty is attempting any type of mapping of the fully anti-symmetric 3N dimensional wave function into three dimensions (see above discussion of normalization). But, if it were possible, what would be the meaning of the scaler value, at every point in real space? Could this value be assigned any meaning? It cannot be a probability. For example, if it were the sum of the probabilities of various electrons occupying a position in space, then the normalized value of the wave function would equal the total number of electrons, yet as discussed above it is always one. If it were the product of probabilities, the normalized value would be less than one. Indeed, the normalized probability of any one electron occupying a particular position is less than one. Clearly, if the individual electron probabilities are normalized, the products of position occupancy will be lower than that of any individual electrons, hence the integrated over space (normalized) value of their products would be less than one.

The above arguments lead to the following conclusion: The multi-electron wave function has no physical meaning. It is merely a mathematical construct to be employed for computing a very limited number of measurable values, as described below. It is interesting to note in this regard that even the most thorough discussions of the physical meaning of quantum theory only discuss the one electron case (13). The argument that wave functions relate to probability are plausible in the one electron case. It is not plausible in the multi-electron case.

The above interpretation shows that in essence there is no information about single electrons available in multi-electron SQQM wave functions. Thus, the multi-electron wave function can be considered a description of a group 'average' behavior, or perhaps is better understood as the behavior of one big electron with a total charge equal to that of the number of electrons in the wave function. It is also clear that the single energy value that can be calculated from the multi-electron wave function is not that of a single electron. It is the sum of the energy of all the electrons. The average radius is that for all the electrons, etc.

This non-intuitive nature of SQQM, the remarkable limitations of the theory are widely understood. Careful physicists understand that the energy values computed using SQQM are only total system values. Thus, for example, in the NIST tables the excitation energies of helium are listed relative to the ground state (23). They are not listed as absolute values, nor is a value provided for either electron independently. Also, in careful computations of the helium energy, no single electron values are provided for either the ground state electrons, nor the 'excited electron'. Only a total value, a single energy, is provided for the system of electrons (24). SQQM, as applied to multi-electron systems, is not consistent with the notion of multiple energy states, one for each electron in the atom.

The failure to provide energy levels of electrons in multi-electron atoms is not only non-intuitive, it is inconsistent with the correlation of one-electron SQQM computations with measured ionization energy for one electron systems (H, He+, Li++, etc.). It is clear that in SQQM that if one were to provide the electron energies for multi-electron systems, they could not be equal to the measured ionization energies, or the 'perpetual motion' conundrums expressed in Equations 1-5 would apply to SQQM just as they do for DQM. The term 'precise energy level' as applied to any electron in an SQQM world is meaningless. All energy level diagrams for electrons, molecules, etc. are meaningless in this version of quantum. And exactly where is phase space?

This analysis of the 'meaning' of a multi-electron wave function leads to an additional puzzle regarding physical understanding of the Pauli Exclusion Principle. This puzzle can be expressed in terms of a three electron atom: If electrons are indistinguishable AND only two can occupy a particular orbital (with opposite and paired spins), which two? In other words, as applied to SQQM the PEP as generally understood is not meaningful. In SQQM there are no independent identifiable electrons, there is only one orbital, one energy, one average radius, for any number of electrons. It doesn't make sense to present an argument regarding how many independent electrons occupy an orbital.  They all occupy the one orbital.  In sum, the PEP as applied to SQQM is a meaningless string of words.  Discussing PEP in relation to SQQM is as meaningful as a discussion of the shape of red or the temperature of a triangle.

What is the value, therefore, of the multi-electron wave function when correctly determined according to the true theory?  Answer: It can be used to compute 'expectation values' for the entire electron collection.  That is, it can produce an expectation value for energy and radius and in a manner of speaking, spin.  But these values are for the 'system as a whole'.  There is absolutely no means to separate electrons and speak of the energy, the orbital, etc. of this or that electron.  After all, there is only one Hamiltonian and it produces one wave function, and that yields one energy.

The total energy for all the electrons, computed using full scale SQQM, can have predictive value.  In the case of helium, this methodology correctly yields a total electron energy equal to the sum of the two measured ionization energies, approximately -79 eV.  Can this be interpreted to mean that each electron has an energy of -39.5 eV? No. To repeat: there is no information about individual electrons that can be extracted from an application of SQQM theory to a multi-electron system.  So, is this result really consistent with the known ionization energies?  Yes, but a second calculation must be performed.  The ionization energy is defined to be the difference between the system at the beginning of ionization and at the end of ionization, that is between a system of N electrons and one with N-1 electrons (more below).  Hence, it is always necessary to make two calculations, one for a system of N electrons and one for a system of N-1 electrons.  In the case of helium the calculation for an N-1 system is really simple as it is for a one electron system and this can be computed using Schrödinger's equation and is approximately -54 eV.  So, the approximate ionization energy is computed to be:

$$E_{ion}= -54 \text{ eV} - (-79\text{eV}) = 25 \text{ eV} \tag{10}$$

in agreement with experiment.

The fact that the SQQM model successfully predicts ionization energies does not make it a scientifically acceptable theory.  Can it produce the multiple ionization energies measured with x-ray photoelectron spectroscopy (XPS) etc?  Not directly, as the only energy value it provides is for the total system of electrons and since it is organized to indicate the electrons in an atom are indistinguishable.   No energy for individual electrons can be extracted from this formalism by its very structure.

THE HYBRID MODELS: <u>DESCRIPTIVE/QUANTITATIVE QUANTUM MODELS (D/Q)</u>-

As DQM has no quantitative content (e.g. can't close an energy balance) and as SQQM cannot even begin to explain many well known physical phenomenon (e.g. multiple ionization energies in multi-electron systems) there is a natural tendency to invent models that are a bit of both.  Spectroscopy clearly shows that atoms have electrons in distinct energy levels.   Hence, scientists search relentlessly for a quantum theory in which energy level diagrams are meaningful, a world in which measured ionization values agree with computed energy levels and a totally three dimensional world. Hybrid models are born from the fact that neither DQM nor SQQM comes close to matching spectroscopic reality.

Simple or naïve models are not satisfactory. For example, a first effort to 'combine' these two theories might be to 'assume' that distinct electrons occupy the 'single particle' wave functions that are combined to create the fully antisymmetric wave function required by SQQM (e.g. Equation 7) and that the overall wave function is just a means to express this mathematically. This can easily be shown to be wrong. If this were true, one could find the energy for each of these states, add the energies and get the same energy as for the fully anti-symmetric multi-electron wave function. It is clear that this is not true just by examining the expressions derived from the multiplications of the wave function(s) and their complex conjugates. There is no overlap in the expressions for the terms in the energy expression from the assumption that the electrons are in independent 'single particle' orbitals, and those for the true result for the full multi-electron wave function. In the 'naïve' one electron 'interpretation':

$$\text{Etot} = \sum_i \langle \phi_1(A) | H_i | \phi_1(A) \rangle \tag{11}$$

where the Hamiltonians in the above expression are some type of 'effective' Hamiltonians, and the sum is over the total number of electrons in the system. In the theoretically correct case (SQQM), using the two electron example for simplicity, the total energy is computed from this expression:

$$E_{Tot} = \frac{1}{2} < [\phi_1(A)\phi_2(B) + \phi_1(B)\phi_2(A)] | H | [\phi_1(A)\phi_2(B) + \phi_1(B)\phi_2(A)] > \tag{12}$$

There is no overlap of any of the terms in Eq. 11 and Eq. 12. Moreover, simply using single particle wave functions derived from perturbation theory it is easy to show that for helium Eq. 11 yields about 70% of the measured total energy whereas Eq. 12 yields better than 95% (25). The low energy computed on the basis of the assumption of independent electrons is one demonstration of the quantitative fallacy of this approach.

Note, single particle wave functions, derived from perturbation theory, lead to notions like 'effective nuclear charge', and 'look' like 'hydrogen orbitals', but from nuclei with reduced 'effective nuclear charge'. However, other 'basis' functions can be used that are not hydrogen-like orbitals. In fact, the 'best' basis set of orbitals by definition is the set that produces the lowest total system energy, not the set composed of hydrogen-like orbitals (25,26). This leads to other questions: If the orbitals don't even resemble hydrogen-like orbitals, on what basis does one still speak of 'orbital angular momentum'? And in the absence of clear values for orbital angular momentum, on what basis do selection rules exist? In general, selection rules for transitions are rationalized on the basis that only transitions that conserve angular momentum are permitted.

The naïve approach outlined above does not work, but it does indicate some necessary elements of any 'theory' that produces both the 'descriptive' elements of DQM and the quantitative elements of SQQM. One element is the requirement that there be a set of Operator Equations, one for each 'independent' electron in the multi-electron system. Solving a set of these equations leads to 'eigenfunctions' and 'eigenvalues' the former are then associated with orbitals and the latter with orbital energy levels. In fact, a significant fraction of all the effort of quantum theorists is focused on defining Hamiltonian-like operators, that is something like the $H_i$ of Eq. 11, as well as the best single electron wave functions (basis set) to use in these operator based quantitative quantum theories.

These methods do improve the performance for predicting energy that match observed ionization energies in some cases, although as illustrated below for the illustrative example selected for this essay, two electron system ionization energies, the computed energy values leave the conundrum of the unclosed energy balance unresolved. Or, if they are resolved, as is sometimes the case for models of molecular orbital energies, the computed energy levels and the measured ionization energies don't match. For example, in those cases in which relaxation is accounted for the computed model energy levels are always lower (more negative) than the measured ionization energies because the 'relaxation energy' is assumed to assist the ionization process.

This inability of the Hartree-Fock and other *ab initio* methods (which all contain *ad hoc* hypotheses) to properly account for relaxation, to obtain energy level sequences in agreement with known sequences, etc. is well known . Perhaps the most famous critics are J.C. Slater (Slater determinant) and his students (27-29). To quote KH Johnson (27): 'In *ab initio* ..methods…of quantum chemistry.ionization energies…usually calculated from Koopman's approximation. …Unfortunately, this assumption ignore the orbital relaxation that generally accompanies ionization. … The relaxation...large in many systems so …Koopmans' lead to inaccurate electron binding energies…. to account for orbital relaxation with the framework of Hartree Fock…. carry out…energy calculations for the neutral system and the ion, (this) requires one to subtract two very large total energies to obtain a small energy difference."

These workers clearly understood that the *ab initio* methods are merely computational methods, not fundamental quantum theory. Hence, they felt empowered to create a competing computational method with no pretenses. The so-called SCF-Xα method employees three 'average' potentials, each within a, somewhat arbitrary, symmetric shaped regions. In the end the symmetric solutions generated in each region are forced to 'match' at region boundaries. It completely abandons the Hamiltonian, etc. However, it often provides energy levels closer to measured ionization energies, (in the correct order) than do any of the so called *ab initio* theories. Still, the SCF-Xα method of computing ionization energy also seems arbitrary. It is based on empirical studies showing that removing half an electron (creating a 'transition state') yields something close to the measured ionization energy. Johnson also points out that the Hartree Fock approach has other failures including giving all excited states positive energy such that there is 'no simple quantitative relationship to the final state of optical excitation.' And he points out that there is no correlation between the order of the orbital energy levels and those observed.

And there is the related issue of computation of excited state energies in those cases for which relaxation energy is considered. For example, atomic helium (notably there is no work in the literature of helium in which energy levels are computed consistent with an accounting of relaxation energy) clearly should have a 'computed' ground state for the two electrons of -39.5, the only value consistent with energy conservation and the PEP (see Figure 2). This energy is necessarily far lower than the measured ionization energy of approximately -24.5 eV in order to account (find a role for) the energy released when the 'unexcited' electron falls to its proper level in He+. As we discuss below, all earlier computations of helium report only the total system energy, ignoring the fact that there computations directly indicate that the two electrons in the ground state must have an energy of -39.5 eV. Another issue: the starting point for computing energy levels using a Hartree Fock approach during excitation. First, there is the need to recognize that two energy levels must be computed. That is, both the excited electron and the 'unexcited' electron levels must be computed. But which model should be applied: Non-Interacting, Initial Drop, Staged Drop (see Figure 2)? And is it not possible to measure these levels. There is always a presumption involved in matching measured values (e.g. -24.5) to very different computed values (e.g. -39.5). Models that cannot be tested experimentally verge on metaphysics.

None of the *ab intio* or other computational methods are to be confused with SQQM. They do not create a single wave function with total symmetry as required by the theory. Instead, employing a variety of assumptions, *ab initio* methods treat a multi-electron system as composed of individual electrons with distinct energies. Also, the equations solved are not Hamiltonian equations. There are sets of operator equations, but these are not Hamiltonians. Moreover, there is one operator equation for each electron, not a single operator equation for all the electrons. This leads to a set of 'orbitals' each with a distinct energy, and in the final analysis the 'orbitals' generated are presumed (by adoption of the PEP) to be doubly occupied by spin paired electrons and the eigenvalues are presumed to be the energies of these electrons. Also, the *ab initio* models don't employ 'phase space'. That is, the 'single electron' Hamiltonian methodologies are a means to create a middle, 'quantitative', ground between DQM and SQQM. To do this they must be neither.

In addition to the earlier described 'relaxation energy' problem there are many other fundamental problems with attempting to make a theory that is somewhat in the middle of SQQM and DQM. For example, interaction energies are determined in these 'one electron' methods employing an inconsistent set of physical interpretations. For example, in the Hartree-Fock method (26) the operator equations (not Hamiltonians!) include two types of integrals that both represent the energetics of interactions between 'distinct electrons': Coulombic repulsion integrals that require the electrons to be independent entities with charge distributions that are 'decreed' to be fully described by 'single particle wave function x single particle wave function[*]', and a second type of repulsive integral for which the underlying basis is 'correlated motion' of point particles.

Detailed examples of the use of the Hartree Fock method to obtain the (real space!) 'orbitals' and 'energy levels' of helium are fully developed. The final form of the equations that must be solved are worth contemplating as even for this simple system it shows all of the deviations of this approach from true SQQM. There are two operator equations, neither Hamiltonian in form, and not one.   The probability distribution is treated as a charge distribution.  Two orbital equations, not a single wave function, are produced.  The two electrons are clearly distinct in character. The two Hartree-Fock equations that must be solved are:

$$hu + Z^{-1}Ju - Z^{-1}Kv = \varepsilon_1 u \,;\; hv + Z^{-1}Jv - Z^{-1}Ku = \varepsilon_2 \qquad (13)$$

where u and v are normalized, orthogonal functions, h is the standard Hamiltonian for atomic hydrogen, and J and K are the coulomb and exchange operators, respectively, defined by:

$$Ju(1) = <v(2)|r_{12}^{-1}|v(2)> \qquad (14)$$

and

$$Ku(1) = <v(2)|r_{12}^{-1}|u(2)> v(1) \qquad (15)$$

and Z is a 'perturbation parameter'.

There are clearly at least three major inconsistencies with SQQM inherent in this approach:  i) The electrons are treated as independent (against the basic assumption of SQQM) particles. ii) These independent particles repel each other both as charge distributions and as point particles. And iii), in order to compute the magnitude of both types of repulsive interactions the wave functions, which in SQQM theory are probability distributions, are treated as charge distributions.  Moreover, these 'charge distributions' have no 'self-interaction' energy, and magnetic interaction energy is not considered.

A fourth 'inconsistency' regarding the use of the Hartree Fock method applies specifically to helium.  For other closed shell atoms there is a long history of computing all the 'energy levels' ( 30-32).  The method historically used to compute the energy levels of the electrons as well as the total energy of all the electrons, explicitly assuming the PEP.  Indeed, this can be readily seen in the equation employed to compute the total energy of electrons in a closed shell system:

$$E_{tot} = 2\sum_{i=1}^{N} \varepsilon_i + \sum_{i,j}^{N} (2J_{ij} + K_{ij}) \qquad (16)$$

where N is the number of doubly occupied orbitals (i.e. 2N is the number of electrons). The first term are the diagonal elements of the final, fully diagonalized, set of equations. The second term effectively makes certain there is proper counting (e.g. no double counting) of electron-electron coloumb and correlation energies.   Equation 16 clearly shows that for helium the H-F method leads to the conclusion that in the ground state the energy of each electron has an energy of one-half the total energy, or each electron is bound at about  -39.5 eV.  Interestingly, in those relatively recent cases in which the H-F method has been applied to helium, only the total energy of the ground state (very close to the measured value) is explicitly reported (33,34).  (Note: for systems of more than two electrons, Koopmans' theory is generally employed to obtain final orbital energies. As shown later this clearly does not work for helium.)

Additional notes about the computation: i) Each Quasi-Hamiltonian operator involves integrations that depend on precise expressions for all the other one electron wave functions. Thus, computations are complex as they must be done iteratively until the final wave function set employed in computing the Hamiltonians is nearly identical to the 'estimated' set of the previous iteration. ii) Full wave functions developed from the set of individual wave functions are not written in a symmetric form, despite the requirements of SQQM (22). iii) There is no term in SQQM Hamiltonians (Eq. 6), nor any observable force, that corresponds to the 'correlation' integrals found in all D/Q models.

D/Q type computational methodologies are so often presented as 'quantum theory' that it is easy to get confused and begin to 'buy' the notion that they are fundamental theory. In fact, they are often labeled as *ab initio* to suggest that they are based on the most fundamental rules of quantum theory. Yet, they are based on 'multiple operators' (not Hamiltonians), rather than a single Hamiltonian and generate multiple 'orbitals' each with a distinct energy. These features of the D/Q models are contrary to the real SQQM theory. Or perhaps the term '*ab initio*' implies that all forces are considered? In that case why is there no magnetic interaction term? These D/Q methods, which come in innumerable variations, are employed because there is no choice. Indeed, given the fact that the very structure of SQQM precludes the possibility of obtaining multiple ionization energies, and DQM is simply not quantitative, only 'computational methodologies' can provide any numerical information that 'sort of' resembles observation.

Koopmans' theorem is the basis for 'teasing out' individual orbital energies from *ab intio* calculations, including Hartree-Fock. This theory requires the construction of a wave functions from a set of N 'single electron wave functions' (e.g. from Hartree-Fock) from which any single electron wave function has been removed. The energy of the new, N-1, wave function can then be calculated and the energy so calculated is the 'energy level' of the removed electron, or:

$$E_N - E_{N-1} = E_{removed\ orbital} \tag{17}$$

Moreover, Koopman's theory requires that the total energy be the sum of all the orbital energies! This clearly does not apply to our standard example, helium! In the helium case Equation 17 becomes:

$$E_N - E_{N-1} = -79eV - (-54.4eV) \sim -25eV \tag{18}$$

If this is correct, then the two electrons in helium, which according to PEP must be identical, cannot be identical, because if they were, then the total energy of helium would be close to -50 eV, which is clearly NOT the value (-79eV) calculated with SQQM or measured experimentally (2-7). Clearly, employing this method returns us to the list of conundrums, including the perpetual motion conundrum, discussed in the section on DQM, above. Hence, all of the D/Q models, which always include the PEP, are inherently inconsistent with an energy balance.

To be specific, Koopman's theory is not consistent with energy conservation for all two electron systems. As an example; in the case of the helium atom an excellent result would be for the Hartree Fock single electron energy level of both electrons to be approximately equal to the measured ionization energy. That is, an 'excellent' result would be achieved if the Hartree Fock method, or any other D/S computation method, predicted a first ionization energy precisely equal to that measured experimentally. However, if this is the energy computed, then as noted repeatedly in this essay (Equations 1-5), there is an energy balance problem. Precisely the same objections raised repeatedly, including the 'perpetual motion' objection, again indicate that the method provides energy levels not consistent with a closed energy balance, and yields a total energy value for the two electron system totally inconsistent with the total energy obtained using any 'total' wave function computation. In fact it is generally understood that *ab initio* approaches do not yield outer electron energies consistent with measured ionization energies (26). Thus, the performance of any pseudo one-electron method is assessed generally by comparing computed energy level spacings with measured energy level spacing. In contrast, ionization energies are usually only computed using the 'double calculation' method described above.

Are the failures of Koopmans' theory to achieve energy self-consistency for helium unique or the rule? It is the rule, as shown below, demonstrating once again that SQQM really says nothing about individual electrons in multi-electron systems.

It can be shown that an absurd 'free energy', as per the section on DQM, is always produced when Koopmans' hypothesis is employed to 'tease out' the energies of individual orbitals, no matter what *ab initio* technique is employed to derive the full wave equation of the multi-electron system. This is inevitable because it is an experimental fact for atoms that the second ionization energies are always higher than first ionization energies. Thus, the sum of the total energy after ionization is never simply the original N atom energy minus the energy of the ionized/missing electron. It is always far lower than that! The system 'relaxes', or at least does so theoretically, to a lower energy. As our first example, we consider alkali earth elements as presumably the outer shell of these atoms, like helium, is occupied by two 's-electrons'.

For all of the alkali earth elements the second ionization energy is nearly twice the first. Moreover, according to Koopmans' model, the highest energy, easiest to ionize, electrons are both in identical 's-orbitals' prior to ionization. Assume one of these electrons is ionized. Measurement of the second ionization energy clearly shows that the s-electron that was not ionized 'fell' in energy by almost 100%! This leads to the same set of questions we encountered in the DQM section. Where does the energy 'go' when the s-electron that is not ionized falls to a lower state? Can't we capture that energy and thereby solve the energy crisis? From whence comes the energy to bring that electron 'back' to its atomic state during electron capture? After all, if a precise amount of energy is required to ionize an atom, then exactly the same amount should be released when the atom recaptures the ionized electron and returns to its original configuration. If more energy is released, then there is a net energy increase in the universe, if less energy is released, then for the universe as a whole energy is being destroyed! In fact, microscopic reversibility is observed. Happily, in the real world energy is not lost or gained during ionization or electron capture.

It is clear according to Koopmans' model, that the net energy of an electron and an alkali metal ion is lower than that of the atomic alkali metal. Hence, this model leads to the absurd conclusion that alkali metal atoms should spontaneously ionize. In fact, a brief look at ionization energies shows that for ALL atomic species each ionization energy is larger than the preceeding one. There are never two ionization energies in a row for any atom that are the same. Where does the energy go each time the electrons 'relax' following ionization? Where do 'relaxed electrons' find the energy to regain their old energy status during subsequent electron re-attachment, a process known to release energy?

It is possible, using the same approach suggested above for DQM (see Figure 2), to generate a multi-wavefunction D/Q 'solution' consistent with energy conservation and the PEP. This is done, in an empirical fashion, by assuming that 'relaxation energy' is part of the energy that contributes the ionization process. This is often done in computations of molecular orbital energies, and the result is that the computed energy levels do not match any measured values. For example, the binding energy computed using this approach is necessarily greater (see Figure 2) than the measured ionization energies. There are at least four major conceptual problems with this approach. First, the 'boost energy' from relaxation is always made assuming that the second ionization energy matches the true energy state of the relaxed electron after ionization. Yet, reflection shows, this energy state cannot be correct, because this level in turn must be corrected for the 'relaxation' taking place during the next ionization process. In short the nesting of 'energy boost' corrections is not considered. Moreover, since all the electrons are supposed to be 'relaxing' two complete calculations of all energy levels, one for the 'before' ionization calculation and one for post-ionization are required. As pointed out by Slater and Johnson, the (molecular) calculations often predict the re-ordering of energy levels (27-29). Second, the energy levels computed never match any measured values. Just as '-39.5 eV' for helium (see Figure 2, compare with Figure 1) does not match any measured value, the proposed energy levels never correspond to anything measured. Too many assumptions, combined with no direct substantiating measurements, puts a theory in the realm of metaphysics. Put another way: the entire basis for accepting SQQM in the first place is the clear correlation for hydrogen between spectroscopic results and the predictions of theory. The scientific method demands such a clear correlation. This correlation does not exist for D/Q theory, and 'faith' is not a scientific commodity. Third, there appears to be no need to consider relaxation 'boost' when inner electrons are directly ionized. In this case, the post-ionization 'relaxation energy' can be accounted for by energy emission in the form of photons or electrons. Why is 'boost' only a consideration for the ionization of the outermost electron? Finally, to make the entire complex web of proposed energy levels work with energy conservation and the PEP, invariably all sorts of variable parameters are introduced in the calculations.

Alternative Models- Are there methods other than SQQM that yield solutions closer to the truth? Indeed, in 'methods' such as SCF Xa and DFT, that abandon the 'wave function' mathematics, treats electrons as individual charge distributions (i.e. they have no 'particle like' properties at all), compute interaction energy on the basis that electrons are charge distributions with 'extra repulsion' due to 'correlation', etc. generally yield results closer to reality than SQQM. But, it is not tenable to claim that these are simply modifications of SQQM.  These computational approaches are in fact distinct theories e.g. see above discussion of some features of SCF - X$\alpha$).  If they are sometimes not presented as such, it is likely for reasons of 'political correctness'.  However, as distinct theories they are not proper subjects of this essay.

STANDARD QUANTUM SUMMARY

 A review of the PEP/closed energy balance issue in relation to each of the three quantum model classes serves as a good summary of the above:

1.  There are at least three categories of quantum theory.  They are incompatible and mutually exclusive. For each the 'meaning' of the PEP is different.

2. An analysis of DQM, as it currently is described in the literature, and its relationship to measured ionization energies, in particular helium, clearly shows that the model predicts infinite energy can be obtained via a process of ionization, and subsequent electron capture. No fuel is modified, yet the prediction of the model is that energy is generated.  This is clearly not consistent with energy conservation. Nor is there any source of energy to 're-position' the 'relaxed electron' into its 'original energy level' following electron capture.  And a simple energy balance show the model predicts helium should spontaneously ionize in order to attain minimum energy.

3.  The only means, and this is not done in the literature, to bring an energy balance and PEP into alignment for the DQM model is to assume the relaxation energy 'boosts' the ionization process.  This in turn indicates the occupied energy levels are far 'deeper' than the measured ionization energies.  In particular, for two electron systems, in both DQM and D/Q methods, the ground state energies must be exactly one-half the total energy of the electron system.   For helium this means they are -39.5 eV, about 15 ev less than the measured first ionization energy.   This directly shows Koopmans' model does not apply to two electron systems.  This requirement also brings up a host of questions regarding the levels occupied by both electrons, not just the higher energy electron, during the excitation process (see Figure 2).

 4.  Precisely the same objections listed above for DQM, apply to all D/Q methods.  In all those methods it is assumed that both helium electrons in the ground state are at the same energy.  In fact, as discussed, this objection applies to any (and that is all!) atomic systems in which the second ionization energy is greater than the first.  'Boost' corrections are also possible with this method, but the net result is a model of postulated energy levels that cannot, by the very nature of the computational process, directly match any spectroscopic data.  And one final objection: D/Q models are not based on Shrodinger's equation.  The various versions of these 'hybrid' theories, each with individual orbitals for each electron, and distributed charge interactions, are 'acceptable' only as computationally tractable 'approximations', not as correct quantum mechanical descriptions of the world.

 5.  It is meaningless to discuss the PEP in relation to SQQM.  All the electrons are indistinguishable.  There is only ONE wave function.  There is only <u>one orbital</u>, and all the indistinguishable electrons are in it.  How does one determine which electrons are in a particular orbital when there are no particular orbitals?

    In the end, it is clear that only SQQM can close the energy balance for helium, hence, despite the fact that probably 99% of all computations are done using a D/Q method, only SQQM is worth discussing.  Yet, SQQM is a very limited model, capable only of computing total properties for multi-electron systems such as total energy, average

radius, etc. The mathematics of SQQM essentially treats a multi-electron system as ONE BIG ELECTRON with one orbital, one energy, one average radius, etc. Absurd conclusions are inevitable when the method is misrepresented as a means to provide information about individual electrons in the multi-electron system. In other words, SQQM is acceptable only for finding the total energy of a system, and the change in the total energy of a system once one electron (in the highest energy 'orbital'?) is removed. The truth is the SQQM model cannot be employed or 'manipulated' to provide data regarding energy states, etc. for individual electrons in a multi-electron system. This inability to yield individual electron energies is not a minor issue. It means the theory is inconsistent with some of the clearest facts in physics such as well-defined, multiple ionization energies for atoms.

Finally, multi-electron SQQM is not acceptable as physics at all because it is not a 'physical' model, unless 3N phase space is to be considered physical. It is only a mathematical construct. In fact, it is an inconsistent mathematical construct.

NEW MODEL

Where do we go from here? Since it was demonstrated that current literature versions of DQM, SQQM and D/Q models are completely inconsistent with each other, with the most common observations, spectroscopic data or energy balances in the case of DQM and D/Q, and multiple ionization energies in the case of SQQM, we clearly need a new theory. The new theory should: 1) be consistent with all classical laws of physics at all scales, 2) lead to quantitative computations that close all energy balances, 3) be quantitatively consistent with the observation that in all multi-electron systems (e.g. atoms) distinguishable electrons are found in distinct sharply defined energy levels, 4) provide a clear value of the angular momentum of each electron in the multi-electron system, and 5) provide predictions directly (not indirectly) comparable to measured data. Specifically, we need a paradigm shift. One is offered below. That is, we attempt to explain via one example how the Classical Quantum Mechanics (CQM) theory developed by R. Mills overcomes all of the inherent difficulties of SQQM and DQM. Specifically, via one example, bound two electron systems, we show how a properly modified CQM closes the energy balance and provides first and second ionization energies in direct agreement with measurement. This work also leads to the conclusion, and this is the modification to CQM offered herein, that in order to close energy balances the Pauli Exclusion Principle must be amended.

The improvements of the CQM model over SQQM as applied to two electron systems are as follows: Maxwell's equations (35,36) and Newton's Laws work on all scales. Each electron is an individual particle with its own energy, magnetic moment, angular momentum and size. There is no Schrödinger wave equation. There is no uncertainty principle. There are no variable parameters (i.e. all constants employed below are from NIST). There is no 'phase space'. And the quantized energy levels have energies in complete agreement with ionization energies. However, unlike DQM, this is a quantitative theory. Unlike SQQM this is a quantitative theory that predicts directly measured energy levels, spin, orbital angular momentum, radii of individual electrons. And unlike SQQM the fundamental equations employed to compute energy levels are simple algebraic, Newtonian force balances. Also, there is no difficulty tracking angular momentum.  It is explicit that each electron has a precise angular momentum.  Finally, the CQM model suggests a truly revolutionary conclusion (a conclusion not supported by RM, the developer of CQM (1)):  The Pauli Exclusion Principle is wrong. A new exclusion principle is postulated, based on the notion that electrons are actually physical objects:  *No two electrons can occupy the same space at the same time.* That is, even if the spins are 'opposite' no two electrons can be identical in all other respects.

   CQM of Two Electron Systems- The CQM model, developed previously entirely by Dr. Randell Mills (1,37), includes the revolutionary hypothesis that bound electrons around single nuclei (atoms) are spherical symmetric bubbles of charge, with a very specific mass and very specific charge current pattern.  These bubbles of zero thickness charge ('orbitsphere') are not solutions to a wave equation. (Note:  The orbitspheres can have orbital angular momentum, and in those cases two dimensional waves, solutions to the two dimensional wave equation, are found on the two dimensional orbitsphere surface.  However, for helium and other two electron atoms in the ground state there is no orbital angular momentum on either orbitsphere.  Hence, for the systems discussed below there is no need to 'confuse the issue' by any consideration of 'waves'.)  It is simply postulated that these bubbles will have properties consistent with all the valid scientific observations regarding bound electrons including the right quantized energy levels, the correct magnetic moment (determined with Maxwell's Laws) and the correct g-factor (angular momentum determined using standard mechanics).  Moreover it is postulated that these bubbles of negative charge obey Maxwell's equations and Newton's Laws. Most of the above has already been thoroughly demonstrated and vetted in the reviewed literature (37).  For example, it clearly is shown that these orbitspheres, with the postulated current pattern, do have the correct angular momentum and magnetic moment. It is also clearly shown that they will not radiate.  Although sophisticated arguments are found elsewhere to demonstrate that closed loops of charge in motion will not radiate (37-39), it is clear enough from ordinary experience.  Indeed, it is well known that currents in superconductors, for example in magnets formed in loops, do not radiate.
       Of particular importance in evaluating this model: There is no requirement in physics for a bound electron to obey a wave equation.  All that is required is that the model be consistent with all objective scientific observations.   For example, a model that is consistent with energy conservation and involves no 'wave' is a valid model, whereas a model that waves but does not conserve energy is not a valid model.
       In this paper we do not have the ambition or need to repeat the arguments made for the success of the orbitsphere in producing quantitative results in precise agreement with objective scientific observations about angular momentum, magnetic moment, g-factor etc. of bound electrons. The goal of this paper is simpler:  To show that two electrons, each assumed to be simply a spherical shell of charge of 'zero' thickness (an 'orbitsphere'), that obeys Newton's Laws and Maxwell's equations, will provide a solution to the above energy balance conundrum, while yielding quantum energy levels precisely equal to measured ionization energies.  The reader unfamiliar with the 'proofs' offered in earlier papers demonstrating that the properties of the orbitsphere are consistent with objective scientific observation, is urged consider the arguments below and to avoid the impulse to 'fire from the hip'.  These readers are urged to take the

following principled pledge of the honest skeptic: *I will expend effort studying the CQM model of the bound electron (i.e. orbitsphere), if and only if it can be shown the orbitsphere model provides quantitative predictions for two electron atoms more consistent with objective scientific observations, including energy conservation, than SQQM.* That is, if the reader emerges from a study of the arguments laid out below, convinced that simple force balances applied to the CQM 'orbitsphere' produces predictions of two electron systems completely consistent with all objective scientific observations of two electron systems, then they should retire to detailed study of earlier papers that purport to show that the 'orbitsphere' model of the electron has the properties claimed, particularly: non-radiation, correct angular momentum, and correct magnetic moment.

The legitimacy of efforts to find physical models for electrons/atoms is verified by the fact that others (9-12) have attempted to find physical models of the electron. These earlier efforts failed (13), but that does not imply that the task is impossible. Indeed, one remarkable, new, epiphany is added in the CQM model: the electron is a spherical shell of charge surrounding the nucleus. All prior efforts assumed a planetary structure for bound electrons.

CQM Model of the Helium Atom: In this model the atom is built in two stages from its constituent parts. (As the structure of helium is a 'state property' the path of construction will not impact the outcome.) In the first step a single electron/orbitsphere is 'added' to a nucleus consisting of two protons, and one or two neutrons to create a He+ ion. *This first step, and all related equations, is a virtual replication of work done by Mill's (1).* It is included for clarity. In the second step, a second electron/orbitsphere is added to the ion created in the first step. A helium atom is the product of this second electron addition. (Most of the mathematics of the second step can be found in Mills's work, but the conception of an 'outer' and 'inner' electron in a two electron system, rather than two electrons of the same energy, represents a significant modification of the original CQM model of two electron atoms. (Mills accepts the PEP. This essay argues the PEP cannot be correct because it is not consistent with the simplest energy balance.) For both steps, the arguments below only concern computation of stable final state properties, not the mechanism of the process.

Step 1: The following are postulated regarding the orbitsphere: It is a physical object shaped like a 'soap bubble' of zero thickness that symmetrically surrounds the nucleus, obeys Newton's Laws and Maxwell's Equations, has a constant angular momentum, h-bar. Clearly, it obeys standard orbital mechanics. Given these requirements, the following force balance equation pertains:

$$\frac{mv^2}{r} = \frac{Ze^2}{4\pi\varepsilon_0 r^2} \tag{19}$$

Where r is the unknown radius, v is the velocity of all the mass on the orbitsphere surface, e is the charge on an electron, $\varepsilon_0$ is the permittivity of free space and Z is the number of protons in the nucleus. In essence this is a very simple statement of Newtonian mechanics: The centripetal force must equal the central force, and the central force is the standard form of the electrostatic interaction between opposite charges. Next, we replace 'v' by noting that the angular momentum of all electrons must be h-bar:

$$v = \frac{\hbar}{m_e r} \quad (20)$$

The angular momentum equation correctly reflects the structure of the orbitsphere: a set of infinitely thin rings of charge/mass, each of the same mass and traveling at the same velocity, and each describing a great circle path around the nucleus. As the loops are all great circles, each has the same radius. Equation (20) is the correct form for the angular momentum of an infinitely thin ring (40). These overlapping and crossing rings are woven to produce a vector projected net angular momentum consistent with experimental measurements, as described elsewhere (1,37). It is legitimate to argue that in classic theory spinning mass rings cannot cross. Thus, it may be correct to say: *In CQM one aspect of elementary particle behavior is postulated that in a classical physics sense is 'unphysical'. Specifically it is postulated that physical crossing of mass/charge rings without interference takes place in elementary particles.* Another required postulate of CQM that the scaler sum (not the vector sum) of the angular momentums of all the rings is h-bar:

$$\hbar = vr \sum m_{ring} = m_e vr \quad (21)$$

Determining the vector sum of the angular momentum from the rings is a more complex geometric computation and is carried out elsewhere (1, 37). Substitution of Eq.20 into Eq. 19 yields:

$$r = \frac{4\pi\varepsilon_0 \hbar^2}{Ze^2 m_e} \quad (22)$$

In fact, $m_e$ should be replaced by reduced mass, $m_r$, where m is the mass of the atom:

$$m_r = \frac{m_e m}{m_e + m} \tag{23}$$

even though this is a very small correction.
Given this definition of Bohr radius:

$$a_0 = \frac{4\pi\varepsilon_0 \hbar^2}{e^2 m_e} \tag{24}$$

A very simple formula for the radius of the first electron in any one electron system is obtained:

$$r = \frac{a_0}{Z} \tag{25}$$

This is significant as the energy of any object bound by a central force with inverse square strength can be obtained from standard mechanics once the mass, radius and angular momentum are determined. In these central force systems the kinetic energy (always positive) is equal to one-half the magnitude of the potential energy (always negative). And the total binding energy will be the sum of kinetic and potential energies. Thus, the total binding energy ($E_B$) will equal, in magnitude (not sign), the total kinetic energy ($E_{kin}$):

$$E_B = -E_{kin} = -\frac{1}{2}mv^2 = -\frac{1}{2}\frac{\hbar^2}{mr^2} = -\frac{1}{8}\frac{e^2 Z^2}{\pi\varepsilon_0 a_0} \tag{26}$$

This simple equation can be solved to yield the one electron energy for all one electron atoms simply by changing one parameter, Z- the number of protons (Table I). (Note: it is a simple matter to determine independently the potential and kinetic energy of the orbitsphere, add them together, and obtain the same result.)
    Several comments regarding the above development are of value. For example, terms not included in the energy balance include 'self-interaction'. (No self interaction terms

are included in SQQM either. Instead, the infinite self-interaction of point particles is swept away in a mathematical vortex known as 'renormalization'.) Not including any self-interaction terms in the CQM model is fully justified by Maxwell's Laws. First, as the orbitsphere is a perfect conductor, it has no field inside. Faraday's Law: There is no field inside a closed conductor. Hence, there is no interaction between charge and an internal field. Is there a field inside the charge layer itself? It is clear that the field is discontinuous. Inside the orbitsphere there is no field, outside there is a well defined field, equivalent to that produced from a single charge at the center of the orbitsphere. Where is the discontinuity? It is on/in the charge layer itself. This is in fact a standard postulate of E&M. There is no field generated by the charge in the layer of the charge, hence there is no self-interaction term. Alternatively, symmetry indicates there can be no field in the plane of a sphere. If the charge distribution is symmetric, there can be no preferred direction. The net result is that at any single point, symmetry shows that the fields must cancel. All points on the orbit sphere are electrostatically equivalent, hence the fields cancel at all points.

There are small corrections that lead to even greater accuracy. For example, as the electron is a real object, it has real velocity and thus there should be relativistic corrections. As the number of protons increases, the radius decreases and the velocity increases in order to maintain h-bar of angular momentum. Thus, the relativistic correction is only significant for large Z. This and other small corrections are discussed in detail elsewhere (1).

Table I: Force balance calculations of the ground state energy of Orbitspheres in one electron systems. Comparison with experimental values is excellent.

| ONE ELECTRON SYSTEM | CQM CALCULATED RADII IN UNITS OF BOHR RADII (Eq. 25) | CQM CALCULATED IONIZATION ENERGY, eV (Eq.26) | EXPERIMENTAL IONIZATION ENERGY, eV |
|---|---|---|---|
| H | 1.00 | 13.61 | 13.59 |
| $He^+$ | 0.500 | 54.42 | 54.58 |
| Li$^{2+}$ | 0.333 | 122.45 | 122.45 |
| Be$^{3+}$ | 0.250 | 217.69 | 217.71 |
| B$^{4+}$ | 0.200 | 340.15 | 340.22 |
| C$^{5+}$ | 0.167 | 489.81 | 489.98 |
| N$^{6+}$ | 0.143 | 666.68 | 667.03 |
| O$^{7+}$ | 0.125 | 870.77 | 871.39 |

    Finally, it is interesting to note that the CQM model is just as accurate as SQQM model for the one system to which SQQM can be applied without approximation or mathematical sophistry of any kind: the excited states of one electron systems, e.g. atomic hydrogen. Again, simply algebraic force balances are all that are required. This is discussed elsewhere (1) but will not be pursued here in depth, because it requires introduction of another concept: photons as trapped electromagnetic energy in resonant cavities (i.e. orbitspheres of precise size). In the CQM model, photons do not mysteriously disappear upon capture as they do in standard physics, their energy converted to a higher potential for an electron. Rather, in CQM photons never cease to exist, but they can be trapped in the resonator cavity that is the orbitsphere. In this essay the minimum number of CQM concepts required to produce a resolution to the issue of closing the energy balance are introduced. The reader interested in other topics such as the excited states of one electron atoms is urged to pursue the matter (1,37).

Step 2: In the second step, an atom is formed when a free electron is captured by the ion created in Step 1. As the free electron about to 'attach' approaches the ion it 'feels a force' from a net central field equivalent to that of a single proton. That is, as the single electron on the ion is spherically symmetric, it creates a field equivalent to that of a point electron at its center. (This 'electron generated' field is restricted to the region outside the orbitsphere. As noted earlier it creates no field inside itself.) Thus, the two proton fields and the one electron field sum to create a field, outside the single orbitsphere of the ion, equivalent to that of a single proton. In other words, the orbitsphere field cancels one proton field outside of itself. In addition, the free electron as it approaches the ion feels a force from the magnetic field generated by the moving currents of the orbitsphere already present on the ion.

    A force balance is once again employed to determine the stable radius. (It is a given that the final stable geometry for both electrons will be that of an orbitsphere, as only in that configuration will a bound electron in a central field not radiate.) However, in this case the force balance should have two terms, one for the net one proton central field and one from the magnetic interaction. The form of the electrostatic term in the force balance is standard, as per Eq. 19. However, a new postulate is required for the form of the magnetic interaction. To wit, it is postulated that the magnetic force between any two

'nested' orbitspheres, in which neither has orbital angular momentum, in any atom (of which two electron atoms are a subset), is described with this equation:

$$F_{mag} = \frac{\hbar^2}{r^3 Z m_e} \sqrt{s(s+1)} \qquad (27)$$

where r is the radius of the electron being acted on and s, the spin quantum number, is 1/2. For the purposes of this essay, this expression for magnetic force between two electrons will be taken simply as a postulate. In the full development of CQM, available elsewhere, it is argued that this term is derived from the standard Lorentz force (1). It is a complex derivation. However, the expression given is acceptable as a postulate as it has a simple algebraic form, uses only the most standard constants (Plank's constant, mass of an electron) and has no adjustable parameters. It is left to others to engage in a 'postulate' or 'derived from Lorentz force' debate. For those who must debate, please recall: Postulates are at the heart of all physics. They are permitted.

Clearly the outer electron feels two forces. It feels the magnetic force from the other electron and it feels a net electrostatic force equivalent to one proton. (As noted above the fields at the outer electron arising from the inner electron and one proton, both mathematically represented by point charges at the center of the atom, cancel.) It is worth noting, as discussed elsewhere (1, chapter 9) that the magnetic interaction can be attractive (opposite spin directions) leading to energy reduction (example: the ground state) or repulsive (same spin direction). In sum, this leads to the following force balance for the ground state:

$$F_{tot} = \frac{mv^2}{r} = \frac{e^2}{4\pi\varepsilon_0 r^2}(Z-1) + \frac{\hbar^2}{r^3 Z m_e}\sqrt{s(s+1)} \qquad (28)$$

This can be re-written, using Eq. 19, as:

$$\frac{\hbar^2}{m_e r^3} = \frac{e^2}{4\pi\varepsilon_0 r^2}(Z-1) + \frac{\hbar^2}{r^3 Z m_e}\sqrt{s(s+1)} \qquad (29)$$

Multiply through by $r^3$, and r can be quickly obtained (Bohr radius defined by Eq. 24):

$$r = a_0 [\frac{1}{(Z-1)} - \frac{\sqrt{s(s+1)}}{Z(Z-1)}] \qquad (30)$$

This simple algebraic equation, derived from a straightforward two term force balance, will be shown to provide energies consistent with measurements of ionization energies, to better than two percent accuracy, for all measured two electron atoms. It also yields an absolute value to the size of the ground state for all two electron atoms, a dimension that can be used to predict scattering behavior.

Computing the energy of this second electron is not quite as simple as for the first, because in this case the central forces are not inverse square. The energy must be calculated as the sum of the potential, kinetic, and magnetic energies. The energy balance for the system is:

$$\Delta E_{tot} = \Delta E_{pot} + \Delta E_{kinetic} + \Delta E_{magnetic} \tag{31}$$

where $E_{pot}$ is the electric potential energy. Moreover, it is clear that the change in the magnetic energy is equal to the work done against the magnetic field in moving from r to infinity, and the change in the electric potential energy is equal to the work done against the electric field. Thus, the change in the magnetic energy is determined from an integration of Eq. 27:

$$\Delta E_{magnetic} = \frac{1}{2}\frac{1}{Z}\frac{\hbar^2}{m_e r^2}\sqrt{s(s+1)} \tag{32}$$

Similarly, the change in the electric potential energy can be determined from integration of the electric force term:

$$\Delta E_{Pot} = (Z-1)e^2 / 8\pi\varepsilon_0 r \tag{33}$$

The change in the kinetic energy, of course, is from a positive value to zero. Hence:

$$-\Delta E_{kin} = \frac{1}{2}\frac{\hbar^2}{m_e r^2} \tag{34}$$

Thus the total energy change of the outer electron during ionization, that is the energy input to bring the outer electron from the bound state to the free state is:

$$E_{ion} = \Delta E_{tot} = (Z-1)e^2 / 4\pi\varepsilon_0 r + \frac{1}{2}\frac{1}{Z}\frac{\hbar^2}{m_e r^2}\sqrt{s(s+1)} - \frac{1}{2}\frac{\hbar^2}{m_e r^2} \qquad (35)$$

It should be noted that this equation explicitly shows that the kinetic energy, inherent in any physical object in a stable orbit, is 're-deployed' such that it contributes to the ionization process. Once the simple formula for r (Eq. 30) is substituted in Equation 35 it can be shown:

$$-\tfrac{1}{2}\Delta E_{Pot} = \Delta E_{Kin} + \Delta E_{Mag} \qquad (36)$$

It is interesting to reflect on the qualitative 'mechanics' of this equation. In inverse square force fields, standard orbital mechanics in gravitation fields for example, the magnitude of the potential energy equals twice the magnitude of the kinetic energy. In the presence of an added central force (e.g. magnetic) the velocity required to maintain the object in stable orbit at any particular radius should be larger, hence the magnitude of the kinetic energy larger. Since the changes in kinetic and magnetic energy are opposite in 'sign' during ionization, this qualitative analysis is consistent with Eq. 35. In any event, Eq.36 can be re-written:

$$E_{ion} = \tfrac{1}{2}\Delta E_{Pot} = (Z-1)e^2 / 8\pi\varepsilon_0 r \qquad (37)$$

Can two extremely simple one-dimensional formulas really quantitatively predict all the known ionization energies of two electron atoms? That is, can the simple formula for ionization energy with no adjustable parameters, derived from a simple algebraic energy balance, Eq. 37, and an equally simple solution for the radius, Eq. 30, derived from a one dimensional force balance also with no adjustable parameters, using only NIST values for physical constants really predict all of the known ionization energies for two electron atoms? The answer, as seen in Table II, is a yes. Helium shows the worst agreement, yet is within two percent. As Z increases, the ions become smaller (literally!), and the relative error decreases to a small fraction of one percent.

Table II: Comparison of predicted and measured ionization energies in two electron systems.

| Z | RADIUS BOHR RADIUS UNITS* | CQM COMPUTED IONIZATION ENERGY Calculated**, eV | MEASURED IONIZATION ENERGY From CRC Tables eV | Relative Error*** |
|---|---|---|---|---|
| 2 | 0.5669 | 23.9965 | 24.587 | 0.024016757 |
| 3 | 0.35566 | 76.50902 | 75.64018 | -0.011486488 |
| 4 | 0.26116 | 156.289 | 153.897 | -0.015542863 |
| 5 | 0.2067 | 263.295 | 259.375 | -0.015113253 |
| 6 | 0.17113 | 397.519 | 392.087 | -0.013854068 |
| 7 | 0.14605 | 558.958 | 552.0718 | -0.012473378 |
| 8 | 0.12739 | 747.61 | 739.29 | -0.011254041 |
| 9 | 0.11297 | 963.474 | 953.112 | -0.010871755 |
| 10 | 0.10149 | 1206.552 | 1195.8286 | -0.008967339 |
| 11 | 0.09213 | 1476.84 | 1465.121 | -0.007998657 |
| 12 | 0.08435 | 1774.34 | 1761.8055 | -0.007114576 |
| 13 | 0.0778 | 2099.0537 | 2085.98 | -0.006267414 |
| 14 | 0.07216 | 2450.977 | 2437.63 | -0.0054754 |

| | | | |
|---|---|---|---|
| 150.0673 | | 2830.1124 | 2816.91 -0.004686838 |
| 160.06306 | | 3236.459 | 3223.78 -0.003932961 |
| 170.05932 | | 3670.0178 | 3658.521 -0.003142472 |

 **\* Eqs 24 and 30, \*\*Eq. 37, \*\*\*(Column 4-Column 3)/Column 4**

The values reported and those predicted will never be close enough to satisfy all physicists. Some small corrections are probably required. Several are under consideration. Indeed, in Mills's model a 'magnetic energy of pairing' is computed that dramatically improves the agreement with data (1, chapter 6), however, the origin of this correction is not clear to the present author. Still, the remarkable agreement between measured values and predicted values for such a simple formula cannot be dismissed with a wave of the hand and a reiteration of the cliché: 'We can do better with our present methods that only requires many hours of supercomputer time.' And the cliché ignores the fact that in the first part of this essay it was demonstrated that standard quantum theory is fundamentally wrong. That is, SQQM only computes the energy of both electrons together, and the values of energy computed with DQM and D/Q, always assume the PEP. In fact, there exists not a single D/Q computation for helium in which the only energy consistent with an energy balance, the PEP and the known ionization energies, that is -39.5 eV, is the ground state 'target' energy of the many computational manipulations. In contrast, in the modified CQM model, described above, the 'inner electron' energy is virtually the same as that given in Table I. That is, if it is correct, the modified CQM model presented here *predicts* something truly unique: it *predicts that the PEP is wrong*.

Isn't there some experimental data that *proves* the Pauli Exclusion Principle? Note first that the entire notion of a 'proof' is a misunderstanding in physics. As explained in the section on the scientific method it is possible to disprove a scientific theory. It is not possible to prove it. Also, demonstrating the falsehood of a theory only requires that one issue be unresolved. Thus, the failure of the Pauli Exclusion Principle to be consistent with energy conservation, according to any standard paradigm, or with the predictions of CQM, is reason for doubt. (The only paradigm consistent with PEP and an energy balance for helium is the one introduced at the end of the DQM section of this paper, e.g. both electrons are at -39.5 eV). Infinite evidence of 'consistency' with other types of data (e.g. spectroscopic) cannot overcome the failure of the model to be consistent with energy conservation. Hence, it is clear that any D/Q or DQM model that assumes the PEP must also assume 'boost'.

The reader is urged to avoid the confusion of further consideration of energy level 'relaxation' without direct emission. There is no 'proof' of it. We postulate relaxation is not a real effect, but only one required by DQM and D/Q M versions of standard quantum theory. There is no experimental support for this belief. <u>In fact</u> (not theoretical fiction) the energy levels of the more tightly bound atomic electrons are only slightly perturbed by the ionization of the outermost/highest energy electron. And we note that contrary to standard quantum theory, this observation of fact, and its prediction by CQM theory, does provide consistency with energy conservation in the universe.

Other purported supports for the Pauli Exclusion Principle are equally simple to dispose. For example, all two-electron entities in a zero spin state are Bosons according to CQM theory. That is, as all electrons have h-bar angular momentum, and the two moments in ground state two electron systems are equal and opposite, they 'cancel'. The objects have no net angular momentum. Thus, CQM two electron systems in the ground state are boson, as bosons have no angular momentum. If the magnetic moment of these bosons aligns with an applied field, as required for a species to produce an electron paramagnetic resonance (EPR) effect, this means that there is a magnetic moment. If there is a net magnetic moment, and 'g' is not zero, then there must be angular momentum. Thus a contradiction in classical physics, the basis of the CQM theory, arises for any magnitude magnetic alignment with an applied field for a Boson. Hence, there is no magnetic alignment and no EPR signal for any two electron system in its ground state according to CQM theory. Hence, there is absolutely no basis for arguing that the absence of an EPR signal indicates that the two electrons are 'identical'. Clearly, as is well known experimentally, in those excited states of helium in which neither the angular nor the magnetic moments cancel, an EPR signal will be observed according to CQM theory.

It is also clear from the most recent studies of the ionization of helium that no data contrary to CQM exists (2-7). For example, no mention of a spectroscopic 'emission' during ionization is made in any of the most recent papers. Moreover, neither the observation that there is no time lag between the ionization of the two electrons in a 'double ionization' event or the finding that the total energy required for this process is about 79 eV is contrary to the predictions of CQM theory. On the other hand, the failure to observe a 'missed emission', apparently is contrary to the expectations of standard DQM and D/Q models, as in all literature models both electrons are at about -25 eV in the ground state, but the single remaining electron, after ionization is at an energy nearly 30 eV lower. Moreover, these models apparently predict that there should be a time lag between first and second ionization events. Indeed, one of the electrons must 'fall' to its final state after/during the first ionization event. Only after it occupies its new state can a second ionization, requiring a higher energy, occur.

EXCITED HELIUM

The next obvious issue is excited state helium. How do SQQM and CQM models of the energy levels in excited helium compare? Once again CQM provides simple algebraic formulas that provide close computational matches for every single energy levels of excited states. These algebraic formulas, that only employ well known constants, are based on essentially the same simple energy balance equations provided above for the ground state. Once again, standard quantum provides a few energies that match a few measured values using convoluted, complex mathematics and questionable physics.

<u>CQM Model of Excited States</u>- There are some consistency issues in the CQM approach that are addressed by the modification to CQM provided in this paper. Indeed, to obtain the correct energy levels for excited helium CQM implicitly accepts the postulate put forth here that helium in the ground state contains electrons at very different energies. Specifically, in the published CQM method (1) for excited states of helium two force balance equations are required, one for the 'excited' electron, one for the inner electron. The inner electron force balance is the same as that for a one electron helium ion with a small correction due to the screened magnetic interaction with the outer electron. In every case it is found that the inner electron is at approximately 0.5 Bohr radii with an energy of approximately -54.4 eV as it is for He+ (see Table I). The excited electron, initially at an energy of approximately -24 eV and a radius of approximately 0.567 Bohr, is excited to a higher energy state by the 'capture of a photon', a process described in references 1 and 37.

Is the existence of an electron in helium at -54 eV and a radii of 0.5 Bohr radii in an excited helium atom obtained in the CQM method consistent with the CQM two electron ground state model (1, Ch. 6) presented in an earlier chapter? No. Indeed, for internal consistency of the CQM text both electrons, just prior to excitation, should initially be present at approximately -24 eV and have a radius of 0.567 eV. That is, in the text (1), the two electrons in any 'ground state' have the same energy and the same radius (Chapter 6). Thus, if the model of excitation is to be forced into consistency with the current CQM model of the ground state two electron systems, the excitation model of a later chapter must be modified to explain two simultaneous phenomenon: drop in energy (nearly 30 eV!) and radius of the inner electron, and the simultaneous promotion of the outer electron to an excited state (<25 eV in all cases).

In contrast, the model presented in the present manuscript, that is that *the two electrons of 'ground state helium'*, and all other two electron systems, *are not at the same energy*, resolves this inconsistency. The values for energy and radii computed in the CQM model of excitation are virtually identical to those of the inner unexcited electron in the ground state helium. Hence, it is argued herein that the two energy model presented in this manuscript for ground state helium is consistent with the computations of the excitation model. In contrast the two electron excited state and ground state models in the current version of CQM are not consistent with each other. In sum, the conclusion reached in this paper: Ground state helium has electrons at two distinct energies, is the only conclusion consistent with all data and all CQM computations.

The above paragraph may be better understood if presented in a different manner. To wit: The 'modification' of the CQM model presented in the present manuscript is in fact simply an application of the model of excited states in Mills' book (1) to the one state not considered in that chapter: Two electrons, in their lowest energy state, with anti-parallel spins. (In the Mills book, the first excited state has the two electrons with parallel spins. This is consistent with the notion that the ground state is the state below that in energy, one with an additional attractive magnetic force (negative energy) and concomitantly one less repulsive magnetic force (positive energy). That is precisely the one described for the ground state in this manuscript: two electrons with an attractive magnetic interaction, that is with anti-parallel spins. Apply precisely the same math to electrons with anti-parallel spins as presented in the excited state model of the Mills text, and one obtains the result discussed herein.

In a sense, in the Mills text there are two versions of the ground state for helium, one completed in the text, the one that requires a mysterious 'application' of the PEP (1, Ch. 6). That treatment is unacceptable for two reasons. First, as described above, it creates an inconsistency with an energy balance. Second, that treatment of the ground state requires the two electrons to be at the same place at the same time, leading to the electrostatic repulsive infinity of energy (singularity) that mars all forms of quantum physics of bound electrons, as discussed above. The second, based on two electrons in completely separate orbitals, is only 'completed' in this manuscript (1, Ch. 9). It does not include a 'forced' application of the PEP and is the first version of bound electrons that have independent energies/orbitals ever presented which is completely consistent with an energy balance. Thus, the modification of CQM promoted in this essay is the only means to save CQM from the energy balance conundrum found in any system required to 'perform' according to the PEP.

Returning to the issue of the CQM model of excited states: Given the 'two energy state model', the CQM model of electron excitation is very compelling. Only simple algebraic force balances are required. The only inputs to the algebra equations derived are the quantum numbers of the final excited state! Every measured excitation energy, all 105, is matched to a fraction of a percent accuracy by a computed value. Angular momentum is explicitly preserved. There are no variable parameters. Only four constants are required. It is not surprising that this model, once the arbitrary PEP assumption is dropped, as per the present manuscript, yields a model of ground state energy and spectroscopy for two electron atoms consistent with fundamental physics laws and all observations.

Standard Quantum Model of Excited States- Denizens of the standard physics world are almost universally surprised to discover that standard physics doesn't offer a simple model of excited helium. After all, physics has moved on to dark force, dark matter, quantum entanglement, etc. Surely, helium is well solved. It's the second simplest atom! Yet, at best the models of excited state helium only provide a few (<10) energies that match measured values. The newest plead a shortage of supercomputer time. All employ some sort of variable parameters, generally many, and generally in the electron-electron force ('correlation') term. (Imagine if gravity or electrostatic interactions were allowed a term that could be optimized for each two body interaction. A joke?) And as the best of these are SQQM type computations, there is no conservation of angular momentum, and no relationship between energy and quantum numbers for the excited electron state (20, 41-43).

SUMMARY

Numerous arguments are made in this essay regarding flaws in standard quantum theory as applied to atoms and ions, and an equal number are made in support of a new theory (although limited in this essay to the range of bound electrons), Classical Quantum Mechanics. It is overwhelmingly evident from experimental observation that the inclusion in standard quantum of the Pauli Exclusion Principle dooms those two categories of the theory (DQM and D/Q), unless the 'boost' hypothesis is added, that postulate the existence of independent electrons in atoms to failure on the basis of simple energy balance considerations. For example, it is demonstrated that if two electrons are in an identical state/energy level prior to ionization, and that if the electron of the pair not ionized truly relaxes into lower energy states following ionization, the very process of ionization would generate energy, yes increase the energy in the universe without any concomitant mass loss. This is clearly not true.

The addition of the 'boost' concept allows the PEP and an energy balance to be in harmony. However, if one adds the 'boost' hypothesis there are several new problems. First, it requires that the computed energy levels not match any experimentally measured value. For example, the PEP, with boost, requires energy level for each electron in the ground state of helium to be -39.5 eV, clearly very different from the first ionization energy of -24.5 eV or the second ionization energy of -54.4 eV. Second, major changes in the present theory would be required, such as entirely new computations of the energy state of the un-excited electron during excitation process. Not even a 'talking model' of the energy levels of the 'unexcited' electron currently exists. Indeed, the most basic issues, such as the impact of 'boost' on the energy of excited states relative to the vacuum have never been discussed (see Figure 2).

It is postulated herein that electron energy levels do not relax (not significantly) following ionization. A corollary is that there is no 'boost'. In fact, there is no experimental evidence of either relaxation or boost. Relaxation is only *believed* to occur because it is required by theory in some cases.

The energy balance failures of DQM are generally ignored, or are conflated with process steps, such as change in screening constant following ionization, which are irrelevant to computation of an energy balance (44). Indeed, as energy is a state property only the initial and final state values are required to determine the net amount of energy released and/or adsorbed during a state change. The nature of the particular process of change is not relevant to this computation. In fact, it was easily shown the only means to have harmony between the PEP and an energy balance is to postulate that the energy levels of stable states in atoms do not match any measured experimental values.

The only 'category' of standard quantum modeling for which the PEP does not require the existence of energy levels that don't match ionization energies is SQQM. And that is because for this version of quantum the PEP is meaningless. Discussions regarding the application of PEP to SQQM are equivalent to arguing regarding the shape of red. Moreover, SQQM doesn't provide energy level information for individual electrons. It only provides the total electron system energy.

Although SQQM does not fail the two electron system energy balance test, it is shown that SQQM is not a particularly attractive theory. It is inconsistent with basic spectroscopy (e.g. multiple ionization energies for multi-electron atoms), Maxwell's equations, Newton's Laws, the existence of spin, requires renormalization of self-energy, and fails the self-consistency requirement of valid scientific theories. Moreover, if SQQM provided results in agreement with spectroscopy, the other classes of quantum theory would never have developed. It even requires a new force ('correlation energy'), a corresponding set of optimized variable parameters, and fails to account for obviously existing magnetic interactions between electrons. Multi-electron SQQM wave functions are also clearly not physical. There is no 3-D wave! At best there are waves in 3N (N is the number of electrons) phase space. Exactly where is 3N dimensional phase space?

In contrast, a slightly modified version of a new theory, CQM, developed by R. Mills is consistent with Maxwell's Equations, Newton's Laws and spectroscopy. It is a totally physical theory as it posits that electrons are real particles with very specific real shapes in real space. Magnetic moments arise from moving charge, as per the Maxwell equations. Forces from electrostatic and magnetic interactions are the only forces. There is no 'uncertainty'. However, to make the new theory self-consistent as well as consistent with an energy balance, one modification is required by the modified CQM presented herein: the PEP must be explicitly rejected. Specifically , it is shown that the modified CQM theory predicts very accurately that in two electron systems in the ground state the two electrons have dramatically different energies. That is, using simple one-dimensional, closed form, no variable parameter, Newtonian force balances, the two ionization energies predicted by the theory nearly perfectly match the known first and second ionization energies of these systems (Table II).

Many physicists will proclaim that the PEP has been experimentally validated. Trust, but verify: Ask for the data. Do not accept an interpretation of the data. This author is still waiting.

Occom's Razor.